\documentclass[a4paper,fleqn]{cas-sc}

\usepackage{lineno} 
\usepackage[numbers,sort&compress]{natbib}
\usepackage{graphicx}
\usepackage{multirow}
\usepackage[title]{appendix}
\usepackage{amsmath,amssymb,amsfonts}
\usepackage{amsthm}
\usepackage{mathrsfs}
\usepackage{xcolor}
\usepackage{textcomp}
\usepackage{manyfoot}
\usepackage{booktabs}
\usepackage{algorithm}
\usepackage{algorithmicx}
\usepackage{algpseudocode}
\usepackage{listings} 
\usepackage{subcaption}
\usepackage{float}
\usepackage{hyperref}
\usepackage{siunitx}
\usepackage{color}
\usepackage{placeins}

\usepackage{threeparttable}
\hypersetup{colorlinks=true, linkcolor=black, citecolor=black, urlcolor=blue}
\def\tsc#1{\csdef{#1}{\textsc{\lowercase{#1}}\xspace}}
\tsc{WGM}
\tsc{QE}

\usepackage{natbib}

\begin{document}
\let\WriteBookmarks\relax
\def\floatpagepagefraction{1}
\def\textpagefraction{.001}
\let\printorcid\relax
\shorttitle{Particle swarm optimization fitting of long-range wake potentials for trapped-mode parameter characterization in the HALF storage ring}   
\title[mode = title]{Particle swarm optimization fitting of long-range wake potentials for trapped-mode parameter characterization in the HALF storage ring}

\author{Haiyan Yao}
\author{Tianlong He}
\ead{htlong@ustc.edu.cn}
\cormark[1]
\author{Xiaoyu Liu}
\author{Weiwei Li}
\author{Zhenghe Bai}
\ead{baizhe@ustc.edu.cn}
\cormark[2]

\address{National Synchrotron Radiation Laboratory, University of Science and Technology of China, No. 42, South Cooperative Road, Hefei, 230029, China}
\cortext[1]{Corresponding authors}
\cortext[2]{Corresponding authors}

\begin{abstract}
Accurate extraction of trapped-mode impedance parameters of complex storage ring components is essential for assessing their impact on coupled-bunch instabilities. This paper proposes a parameter extraction method based on particle swarm optimization. By constructing a multi-resonator fitting model, trapped-mode parameters are extracted from partially decayed long-range wake potentials. Benchmark validation using a cylindrical pillbox cavity demonstrates that the proposed method yields results consistent with those obtained from the CST eigenmode solver and the existing differential evolution method for both longitudinal and transverse cases, while significantly reducing the computational cost. The method is further applied to three critical components of the Hefei Advanced Light Facility storage ring, demonstrating its applicability to complex structures.
\end{abstract}


\renewcommand{\thefootnote}{\fnsymbol{footnote}}
\footnotetext[2]{Supported by the National Natural Science Foundation of China (No. 12375324) and the Fundamental Research Funds for the Central Universities (No. WK2310000127)}
\maketitle

\section{Introduction}\label{sec:introduction}
In electron storage rings, when a charged particle beam traverses vacuum chambers and accelerator components, electromagnetic wakefields are excited by finite conductivity and geometric discontinuities. These wakefields act on trailing particles and constitute one of the key mechanisms driving collective beam instabilities. In the frequency domain, such beam--structure interactions are commonly characterized by the coupling impedance~\cite{Métral1,Nagaoka}. For cavity-like structures, the impedance spectrum often exhibits multiple resonance peaks with a high quality factor, known as trapped modes or higher-order modes (HOMs)~\cite{Palumbo}. Owing to their slow electromagnetic energy decay, these modes can generate long-range wakefields once excited by the beam and may accumulate over many revolutions, thereby driving coupled-bunch instabilities. Meanwhile, the electromagnetic fields may also deposit power in vacuum components, causing beam-induced heating, limiting the storable beam current, and, in severe cases, leading to overheating and damage of accelerator components~\cite{Khaldi,Tian,Sangroula,Kononenko}. Therefore, accurate extraction of trapped-mode characteristic parameters is essential for establishing reliable impedance models, performing beam-dynamics studies, and evaluating potential instability and thermal risks, as well as for developing effective mitigation measures.

Trapped-mode parameter extraction currently relies mainly on two types of numerical methods. The first is eigenmode analysis, in which the eigenmode solvers of commercial electromagnetic software packages such as CST~\cite{CST} and HFSS~\cite{HFSS} are used to directly calculate the resonant frequency and quality factor of a structure, while also providing the corresponding field distributions. Eigenmode analysis has been widely applied to the study of HOMs in accelerator cavities and vacuum components~\cite{Salvant,Ericson,Pommerenke,Kim}. However, it typically requires scanning a predefined frequency range for eigenmodes, not all of which are effectively excited under realistic beam conditions. The second type is based on time-domain wakefield simulations, for example using the CST wakefield solver~\cite{CST} or GdfidL~\cite{GdfidL}, followed by Fourier transformation to obtain the impedance spectrum and further analyze narrowband resonant features. This method naturally incorporates the beam excitation process and therefore has clear advantages in studying the coupling between trapped modes and the beam~\cite{Ericson,Marhauser,Faillace}. However, for narrowband resonances, accurately resolving the corresponding resonance peaks generally requires a sufficiently long wakefield simulation, which substantially increases the computational cost~\cite{Khan}.

To address the above issues, optimization-based methods for trapped-mode parameter extraction offer clear advantages in the analysis of narrowband modes. In particular, a series of CERN-related studies has established a development path from impedance extrapolation based on partially decayed wakes to direct fitting of resonator parameters. Shipman~\textit{et al.} first proposed an analytical extrapolation method for longitudinal impedance, in which resonator models are fitted to impedance data derived from truncated wakes. This allows the fully converged longitudinal impedance spectrum to be reconstructed from partially decayed wakefields, thereby significantly reducing the wakefield simulation cost required for narrowband structures~\cite{Shipman}. Subsequently, Joly further developed the resonator formalism for partially decayed wakes and extended it into a more general framework capable of describing longitudinal and transverse impedance, wake functions, and wake potentials, providing a more complete theoretical basis for recovering resonator parameters from finite-length wake data~\cite{Joly1}. Building on this framework, Nielsen introduced the differential evolution (DE) algorithm into resonator fitting and demonstrated the direct recovery of resonator parameters, including the resonant frequency, quality factor, and shunt impedance, from wake potentials or impedance data~\cite{Nielsen}. This DE-based fitting method is also implemented in the open-source IDDEFIX framework~\cite{Joly2}. Nevertheless, when the number of fitting parameters becomes large or the dataset size increases, it still leaves room for improvement in convergence efficiency and computational cost.

This paper presents a particle swarm optimization (PSO)-based fitting method for trapped-mode parameter extraction. The procedure begins with a wakefield simulation of moderate computational cost, from which a partially decayed wake potential and its associated impedance spectrum are obtained. The proposed method treats the resonant frequencies as known inputs to a multi-resonator model and recovers the remaining parameters by applying PSO~\cite{Poli} to a multivariable single-objective fit of the long-range wake potential, enabling efficient extraction of trapped-mode characteristics. This method is applicable not only to the trapped-mode analysis of geometric discontinuities of vacuum components, but also to the HOM-parameter calculation of structures such as accelerating cavities, thereby supporting HOM suppression and structural optimization. To verify its effectiveness, a cylindrical pillbox cavity is first adopted as a benchmark model for both longitudinal and transverse cases. Its applicability to more complex realistic structures is further assessed using three representative components in the Hefei Advanced Light Facility (HALF) storage ring.

The remainder of this paper is organized as follows. Section~2 introduces the basic formulas of the resonator model and the PSO-based fitting method for characteristic parameters. Section~3 presents a benchmark validation of the proposed method. Section~4 demonstrates its application to typical components in the HALF storage ring. Finally, Section~5 summarizes the main conclusions.

\section{Resonator modeling and PSO-based extraction of trapped-mode parameters}
\label{sec:level2}

\subsection{Standard resonator impedance and wake functions}
\label{subsec:resonator_model}
For cavity-like accelerator components, the electromagnetic response is typically characterized by multiple high-$Q$ trapped modes. In the frequency domain, the trapped mode is generally represented by a parallel RLC circuit model. Accordingly, the longitudinal impedance of a single resonator is given by~\cite{Hofmann,Zotter},
\begin{equation}
Z_{\parallel}(\omega)
=
\frac{R_s}
{1 + i Q \left( \frac{\omega}{\omega_r} - \frac{\omega_r}{\omega} \right)},
\label{eq:Zl_resonator}
\end{equation}
where $R_s$ is the shunt impedance, $Q$ is the quality factor, and $\omega_r = 2\pi f_r$ is the resonant angular frequency.
The transverse resonator impedance takes a similar form:
\begin{equation}
Z_{\perp}(\omega)
=
\frac{R_s}
{1 + i Q \left( \frac{\omega}{\omega_r} - \frac{\omega_r}{\omega} \right)}
\frac{\omega_r}{\omega},
\label{eq:Zt_resonator}
\end{equation}
where the additional factor $\omega_r/\omega$ follows from the conventional definition of the transverse impedance through the Panofsky--Wenzel theorem.

The corresponding point-charge longitudinal wake function in the time domain is expressed as \cite{Métral2}
\begin{equation}
w_{\parallel}(t)
=
\omega_r \frac{R_s}{Q}
\exp\!\left(-\frac{\omega_r t}{2Q}\right)
\left[
\cos(\omega_n t)
-
\frac{\omega_r}{2Q\omega_n}\sin(\omega_n t)
\right]
H(t),
\label{eq:wl_point}
\end{equation}
where $H(t)$ is the Heaviside step function and $\omega_n = \omega_r\sqrt{1-\frac{1}{4Q^2}}$ is the detuned resonance frequency of the damped oscillation. For the transverse case, the point-charge wake function can be written as
\begin{equation}
w_{\perp}(t)
=
\omega_r \frac{R_s}{Q}
\exp\!\left(-\frac{\omega_r t}{2Q}\right)
\sin(\omega_n t)\,H(t).
\label{eq:wt_point}
\end{equation}

Equations~(\ref{eq:Zl_resonator})--(\ref{eq:wt_point}) define the standard single-resonator representation used throughout this work. They provide the basic parametrization of a trapped mode in terms of $(f_r,Q,R_s)$, or equivalently $(f_r,Q,R/Q)$.

\subsection{Gaussian-bunch wake potentials and truncated-impedance formulation}
\label{subsec:gaussian_bunch_wake}
For time-domain wakefield solvers such as CST, the driving source is usually modeled as a finite-length bunch, and the computed quantity therefore corresponds to the bunch wake potential. In practical calculations, the bunch profile is generally assumed to be Gaussian. Since the following formulation is written in the time domain, \(\sigma\) denotes the rms bunch duration. The rms spatial bunch length specified in CST is denoted by \(\sigma_z\), and the two are related by \(\sigma=\sigma_z/c\). The normalized bunch distribution is
given by
\begin{equation}
\lambda(t)
=
\frac{1}{\sqrt{2\pi}\sigma}
\exp\!\left(-\frac{t^2}{2\sigma^2}\right).
\label{eq:gaussian_profile}
\end{equation}

The bunch wake potential corresponding to this Gaussian distribution can be expressed as the convolution of the point-charge wake function with the bunch distribution. For a single resonator mode, this convolution admits an analytical expression. The longitudinal bunch wake potential can thus be expressed as
\begin{equation}
\begin{aligned}
W_{\parallel}(t)
&=
\exp\!\left(\frac{\alpha^2-\omega_n^2}{2}\sigma^2\right)
\exp(-\alpha t)
\\
&\quad\times
\Bigg[
C^{\sin}\,
\Im\Bigg(
\exp\!\big(i(\omega_n t-\alpha\omega_n\sigma^2)\big)
\,
\mathrm{erfc}\!\left(
-\frac{t-\alpha\sigma^2+i\omega_n\sigma^2}{\sqrt{2}\sigma}
\right)
\Bigg)
\\
&\qquad+
C^{\cos}\,
\Re\Bigg(
\exp\!\big(i(\omega_n t-\alpha\omega_n\sigma^2)\big)
\,
\mathrm{erfc}\!\left(
-\frac{t-\alpha\sigma^2+i\omega_n\sigma^2}{\sqrt{2}\sigma}
\right)
\Bigg)
\Bigg],
\end{aligned}
\label{eq:Wl_gaussian}
\end{equation}
where $\alpha = \omega_r / (2 Q)$ is the damping factor, the coefficients 
$C^{\sin} = -\, R_s \omega_r^2 / (4 Q^2 \omega_n)$ and 
$C^{\cos} = R_s \omega_r / (2 Q)$ determine the relative contributions of the sine and cosine components of the damped oscillation. Here, $\Im(\cdot)$ and $\Re(\cdot)$ denote the imaginary and real parts of a complex quantity, respectively, $\mathrm{erfc}(\cdot)$ is the complementary error function, and $t$ is the time behind the source charge, with physical interest in the range $t \in [0,+\infty)$. Although the expression appears complicated, it provides an exact analytical description of the Gaussian-bunch wake potential for a single resonator mode. Likewise, the transverse bunch wake potential is
\begin{equation}
\begin{aligned}
W_{\perp}(t)
&=
\frac{R_s}{2Q}\frac{\omega_r^2}{\omega_n}
\exp\!\left(\frac{\alpha^2-\omega_n^2}{2}\sigma^2\right)
\exp(-\alpha t)
\\
&\quad\times
\Im\Bigg[
\exp\!\big(i(\omega_n t-\alpha\omega_n\sigma^2)\big)
\,
\mathrm{erfc}\!\left(
-\frac{t-\alpha\sigma^2+i\omega_n\sigma^2}{\sqrt{2}\sigma}
\right)
\Bigg].
\end{aligned}
\label{eq:Wt_gaussian}
\end{equation}

The corresponding impedance associated with the bunch wake potential can be obtained from the Fourier transform of the wake potential. In practice, the wakefield simulation is performed over a finite length only. If the wake has not fully decayed within the simulation window, direct Fourier transformation of the truncated wake tends to underestimate the true narrowband impedance peak. To address this issue, CERN has developed a method for parameter extraction from partially decayed wakes. The simulated wake is assumed to correspond to the interval $0\le t\le T$, and the resonator model is Fourier-transformed over the same finite interval to construct a truncated impedance model. Taking the longitudinal case as an example~\cite{Nielsen}:

\begin{equation}
Z_{\parallel}^{\mathrm{partial}}(\omega)
=
i\int_{0}^{T}
W_{\parallel}(t)\exp(-i\omega t)\,\mathrm{d}t,
\label{eq:Zpartial_general}
\end{equation}
where $T$ denotes the available wake length in time. The explicit expression of the resulting truncated impedance can be found in Ref.~\cite{Joly1}. For a single resonator, the fitting parameters remain $(f_r,Q,R_s)$, and the fitted parameters can subsequently be substituted back into the standard fully decayed resonator expressions to reconstruct the complete wake potential and impedance. This truncated-impedance formulation provides the basis of the CERN DE extrapolation scheme and serves as an important reference method in the present work.

\subsection{Simplified analytical expressions for long-range wake potentials}
\label{subsec:long_range_wake}
In trapped-mode analysis, high-\(Q\) resonance wake potentials decay slowly, whereas resistive-wall and other low-\(Q\) contributions decay much more rapidly. Fitting the long-range wake potential therefore provides a more direct way to capture trapped-mode characteristics. For a single longitudinal resonator mode, in the limit \(t \gg 0\), the long-range bunch wake potential can be expressed as the product of the point-charge wake function and a bunch form factor~\cite{He1},
\begin{equation}
W_{\parallel}^{\mathrm{long}}(t)
=
-\,F\,w_{\parallel}(t),
\label{eq:Wlong_factorized}
\end{equation}
where $w_{\parallel}(t)$ is given by Eq.~(\ref{eq:wl_point}) and
\begin{equation}
F
=
\int_{-\infty}^{\infty}
\lambda(\tau)
\exp\!\left[
\left(-i+\frac{1}{2Q}\right)\omega_r\tau
\right]
\mathrm{d}\tau,
\label{eq:form_factor}
\end{equation}
where \(F\) is the bunch form factor associated with the resonator mode. Substituting Eq.~(\ref{eq:wl_point}) into Eq.~(\ref{eq:Wlong_factorized}) yields
\begin{equation}
W_{\parallel}^{\mathrm{long}}(t)
=
-\,F\,
\frac{R_s}{Q}\omega_r
\exp\!\left(-\frac{\omega_r t}{2Q}\right)
\left[
\cos(\omega_n t)
-
\frac{\omega_r}{2Q\omega_n}\sin(\omega_n t)
\right],
\label{eq:Wlong_longitudinal}
\end{equation}
and the transverse long-range wake potential can be written as
\begin{equation}
W_{\perp}^{\mathrm{long}}(t)
=
F\,
\frac{R_s}{Q}\omega_r
\exp\!\left(-\frac{\omega_r t}{2Q}\right)
\sin(\omega_n t).
\label{eq:Wlong_transverse}
\end{equation}

Equations~(\ref{eq:Wlong_longitudinal}) and~(\ref{eq:Wlong_transverse}) form the key formulas for the PSO-based fitting of long-range wake potentials in this work.

\subsection{PSO-based fitting of long-range wake potentials}
\label{subsec:pso_fitting}
Using the long-range wake formulas given above, we propose a PSO-based fitting method to extract trapped-mode parameters. Rather than fitting the complete wake or the truncated impedance, as done in the DE-based method described in Section~\ref{subsec:gaussian_bunch_wake}, the present method focuses only on the long-range wake component and incorporates resonant frequencies pre-identified from the impedance spectrum as prior information in the fitting procedure.

For a single resonator mode, the long-range wake potential is characterized by only two unknown parameters, $(A,Q)$, where $A=(R/Q)F\omega_r$. In practical fitting, the search range of \(A\) is determined from the wake amplitude, while the quality factor \(Q\) is constrained within \(10 \le Q \le 10^{5}\) to ensure stable convergence. In contrast, the DE method uses a data-driven strategy to determine parameter bounds. Resonance peaks are first identified from the impedance spectrum, and the corresponding $-3$~dB bandwidths are used to estimate initial values of the quality factor. Based on these estimates, adaptive bounds for $(R_s, Q, f_r)$ are constructed, effectively restricting the search space and improving optimization robustness.

For a structure supporting $N$ resonator modes, the total long-range wake potential is obtained by superposition of the single-mode contributions:
\begin{equation}
W_{\parallel}^{\mathrm{long}}(t) = \sum_{k=1}^{N} W_{\parallel,k}^{\mathrm{long}}(t), \quad
W_{\perp}^{\mathrm{long}}(t) = \sum_{k=1}^{N} W_{\perp,k}^{\mathrm{long}}(t).
\end{equation}

Accordingly, the optimization variables are chosen as $P=(Q_1,A_1,Q_2,A_2,\ldots,Q_N,A_N)$, and the dimensionality of the problem is therefore $2N$. Meanwhile, the sum of squared errors is adopted as the objective function to measure the fitting error:
\begin{equation}
\mathcal{J}(P)=
\sum_{j=1}^{L}
\left|
W_{\mathrm{model}}(t_j;P)-W_{\mathrm{ref}}(t_j)
\right|^2 ,
\label{eq:SSE}
\end{equation}
where $W_{\mathrm{ref}}$ denotes the long-range wake potential obtained from the wakefield solver, and $L$ is the number of sampling points. The PSO algorithm iteratively updates the particle positions to minimize the objective function, thereby determining the optimal long-range resonator parameters. After optimization, the physical resonator parameters are recovered as $(R/Q) = A/(F\omega_r)$ and $R_s = Q\,(R/Q)$.

\begin{figure}[htbp]
    \centering
    \begin{subfigure}[b]{0.45\textwidth}
        \centering
        \includegraphics[width=\textwidth]{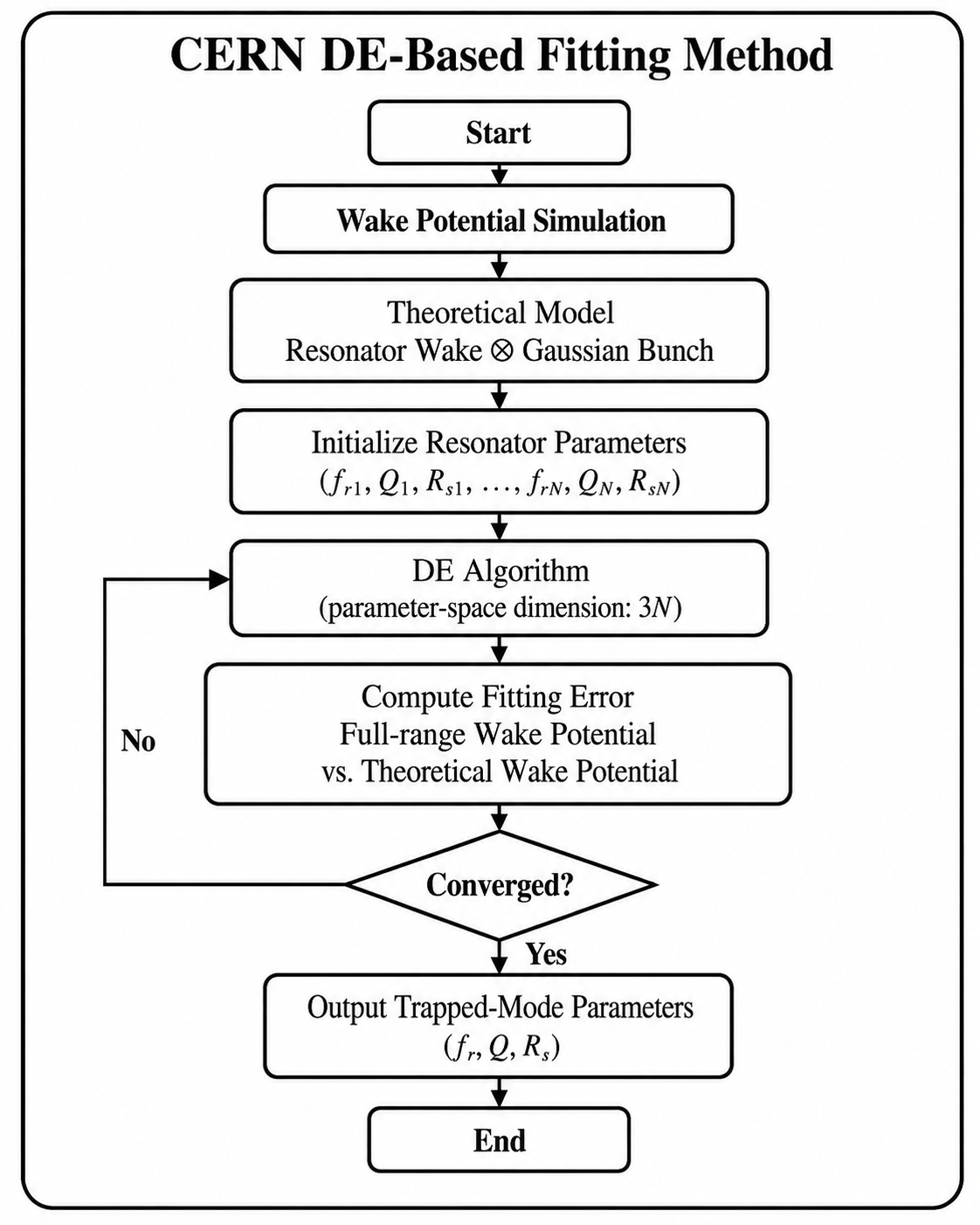}
        \caption{}
        \label{fig:DE_workflow}
    \end{subfigure}
    \hfill
    \begin{subfigure}[b]{0.45\textwidth}
        \centering
        \includegraphics[width=\textwidth]{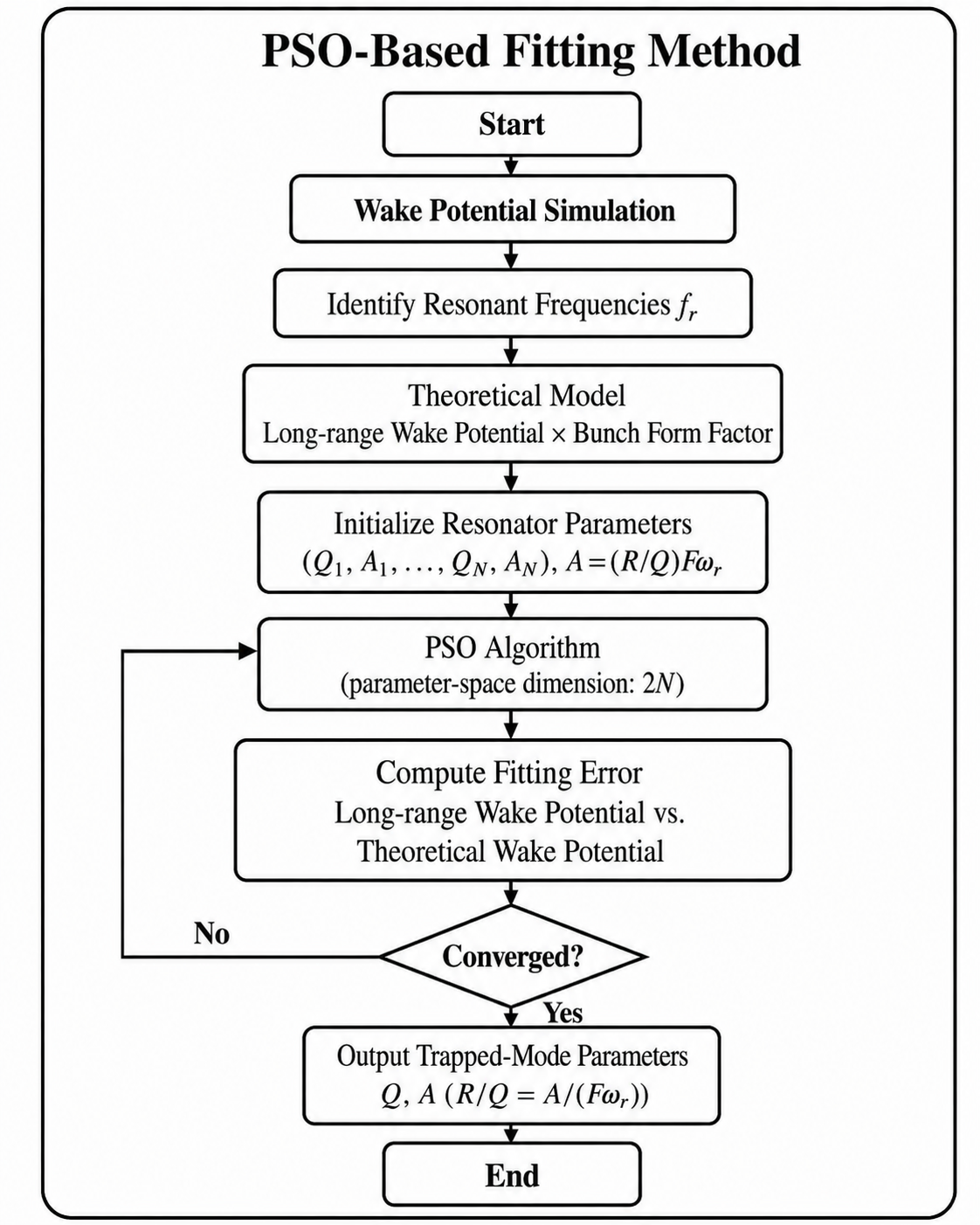}
        \caption{}
        \label{fig:PSO_workflow}
    \end{subfigure}
    \caption{Schematic comparison of the (a) DE workflow and the (b) PSO workflow.}
    \label{fig:DE_PSO_workflow}
\end{figure}

A schematic comparison between the CERN DE workflow and the PSO workflow adopted in this work is shown in Fig.~\ref{fig:DE_PSO_workflow}. In the CERN DE workflow, \((f_r,Q,R_s)\) are used as fitting parameters, and either the full wake data or the truncated impedance is fitted in a \(3N\)-dimensional parameter space. In contrast, the present PSO workflow uses \((A,Q)\) as fitting parameters and fits the long-range wake potential in a \(2N\)-dimensional parameter space by using pre-identified resonant frequencies as prior information.

\section{Benchmark study using a cylindrical pillbox cavity}
\label{sec:level3}
\begin{figure}[htbp]
    \centering
    \includegraphics[width=0.5\textwidth]{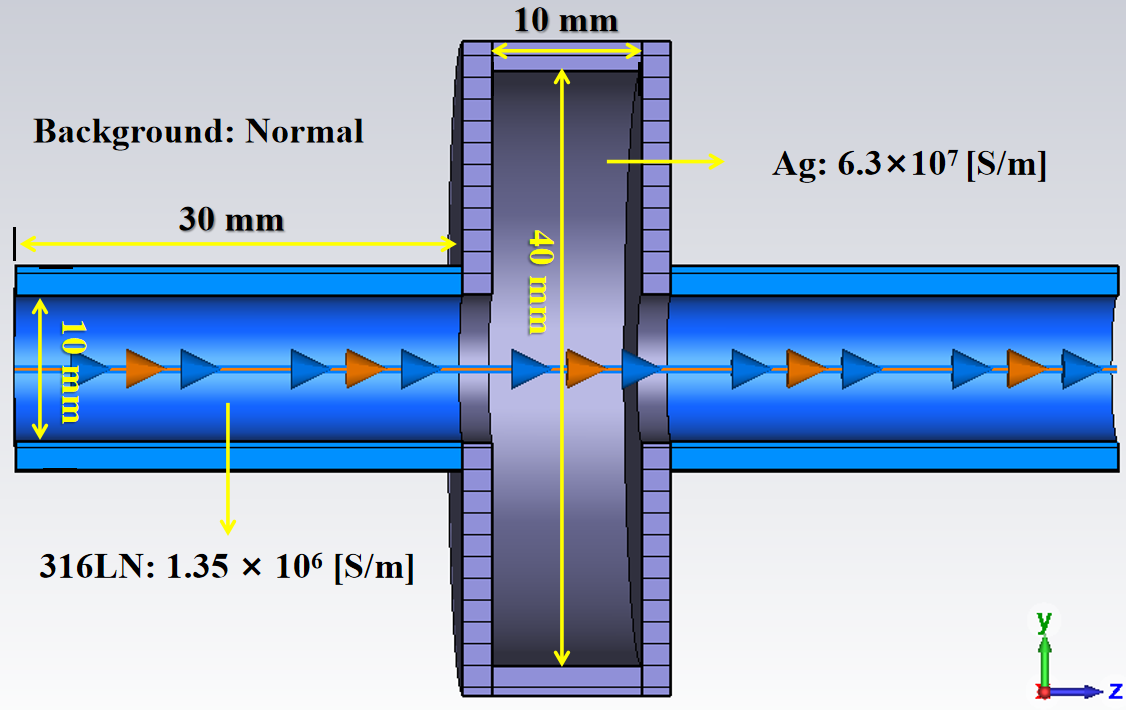}
    \caption{Pillbox model.}
    \label{fig:pillbox_model}
\end{figure}

To evaluate the accuracy and robustness of the PSO method for trapped-mode parameter extraction, a cylindrical pillbox cavity is selected as the benchmark model. Its geometry is simple and its modal spectrum is well separated, without multimode coupling or additional interference. Moreover, the parameters $f_r$ and $R_s$ of the pillbox cavity can be reliably obtained by analytical methods, thereby providing a reference for the quantitative evaluation of the fitting results.
The pillbox cavity used in this study is modeled in CST, as shown in Fig.~\ref{fig:pillbox_model}. The model consists of a central pillbox cavity and two beam pipes on both sides. The central cavity has an axial length of 10~mm, a radius of 20~mm, and a wall thickness of 2~mm. It is made of silver with a conductivity of \(6.3\times10^{7}~\mathrm{S/m}\). The beam pipes have an aperture radius of 5~mm, a wall thickness of 2~mm, and a length of 30~mm on each side. They are made of 316LN stainless steel with a conductivity of \(1.35\times10^{6}~\mathrm{S/m}\). The background material is set to vacuum, and the beam propagates along the \(z\) direction. A quarter-symmetry boundary condition is applied in the simulation. Open boundary conditions are applied in the longitudinal direction, while a perfect-electric-conductor boundary condition is imposed on the radial outer boundary. The bunch length is set to 6~mm, and the time-domain wakefield simulation is performed using the CST wakefield solver.

For validation, CST wakefield simulations are first carried out over a finite wake length to obtain wake-potential data. The PSO method is then applied to fit the wake potential and extract the resonator parameters, which are compared with those obtained from CST eigenmode analysis and the DE-based impedance fitting method. This comparison provides a basis for evaluating the accuracy, consistency, and computational efficiency of the PSO method. The extracted longitudinal and transverse trapped-mode parameters are presented in the following sections.

\subsection{Longitudinal trapped-mode parameter extraction results}
\begin{figure}[htbp]
    \centering
    
    \begin{subfigure}{0.48\linewidth}
        \centering
        \includegraphics[width=\linewidth]{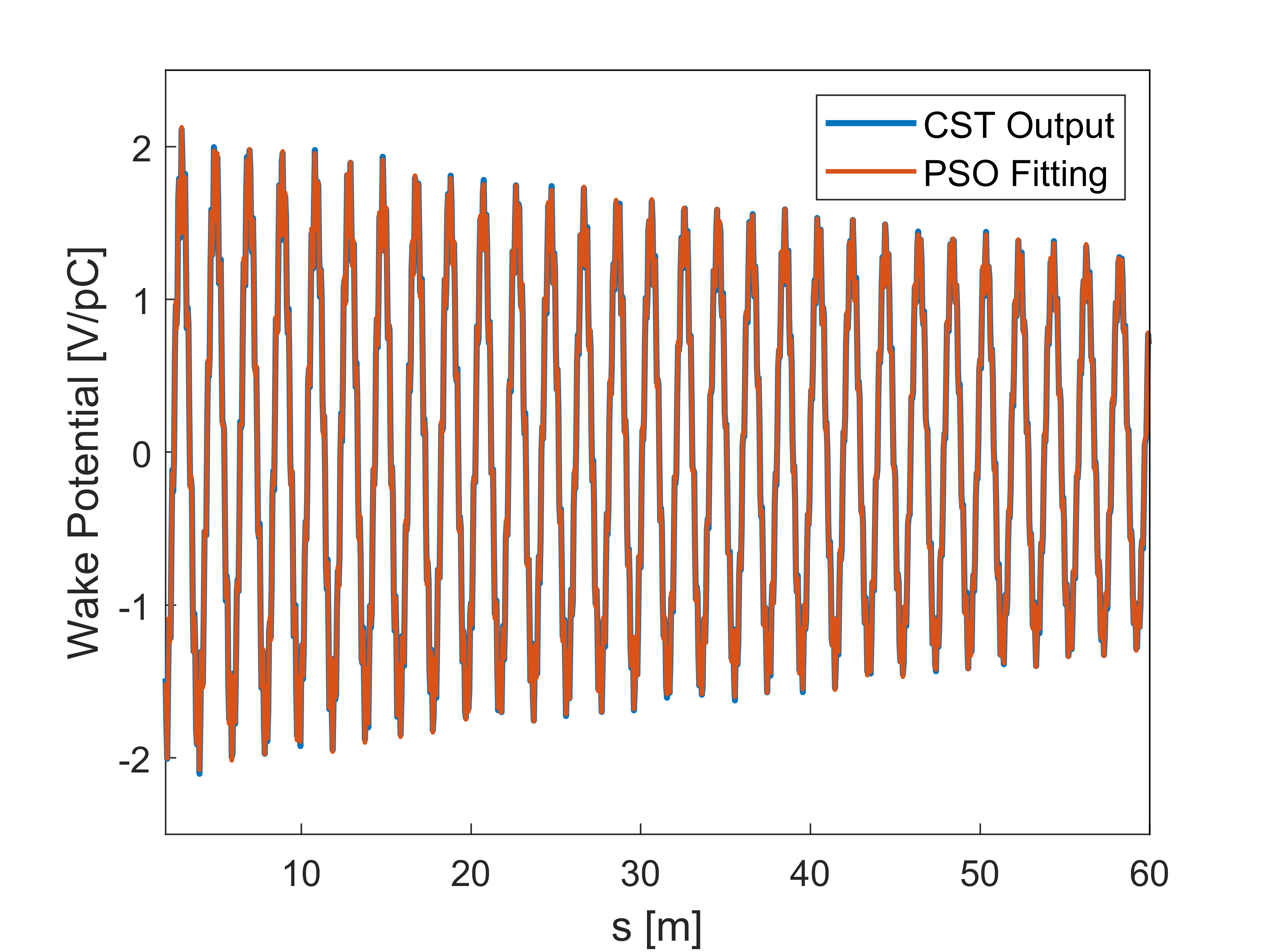}
        \caption{}
        \label{fig:60mZ_wp}
    \end{subfigure}
    \hfill
    \begin{subfigure}{0.48\linewidth}
        \centering
        \includegraphics[width=\linewidth]{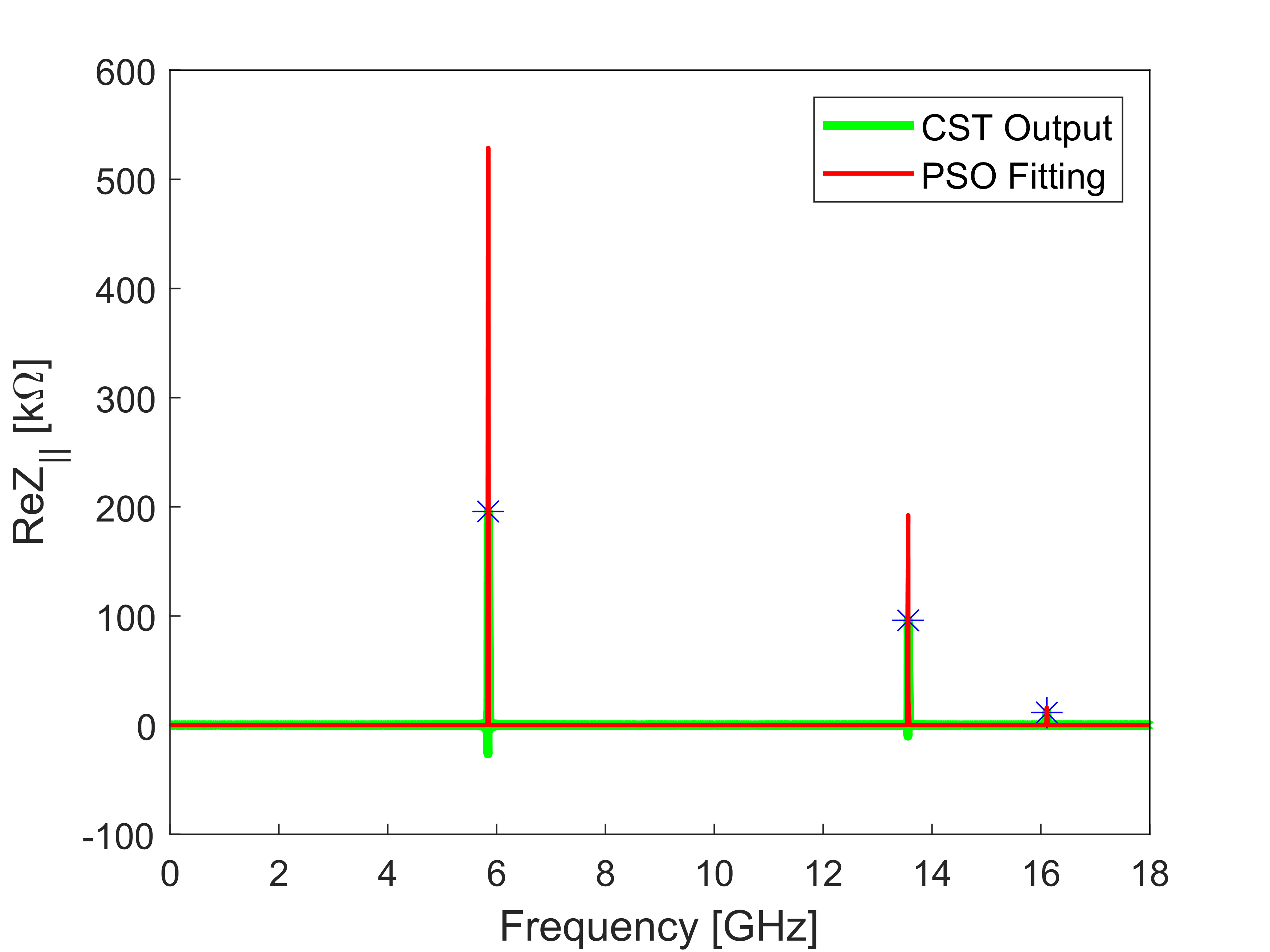}
        \caption{}
        \label{fig:pillbox_60m}
    \end{subfigure}
    
    \caption{PSO-based results for the pillbox cavity: (a) fitted longitudinal wake potential and (b) reconstructed longitudinal impedance. The asterisks denote the impedance amplitudes obtained by Fourier transforming the truncated CST wake potential.}
    \label{fig:pillbox_combined}
\end{figure}

For the longitudinal case, the beam offset in the CST wakefield solver is set to zero, and the wake length is extended to 60~m. The resulting longitudinal wake potential is fitted using the PSO algorithm, and the corresponding longitudinal impedance is reconstructed from the fitted resonator parameters. As shown in Fig.~\ref{fig:pillbox_combined}(a), the fitted wake potential agrees well with the CST simulation result, and the two curves nearly overlap over the full fitting range. However, the wake potential has not fully decayed even at a propagation distance of 60~m. As a result, directly Fourier transforming the truncated CST wake potential underestimates the narrowband impedance amplitudes. This can be seen in Fig.~\ref{fig:pillbox_combined}(b), where the reconstructed longitudinal impedance is compared with the impedance amplitudes obtained from the truncated wake.

\begin{figure}[htbp]
    \centering
    \includegraphics[width=0.6\textwidth]{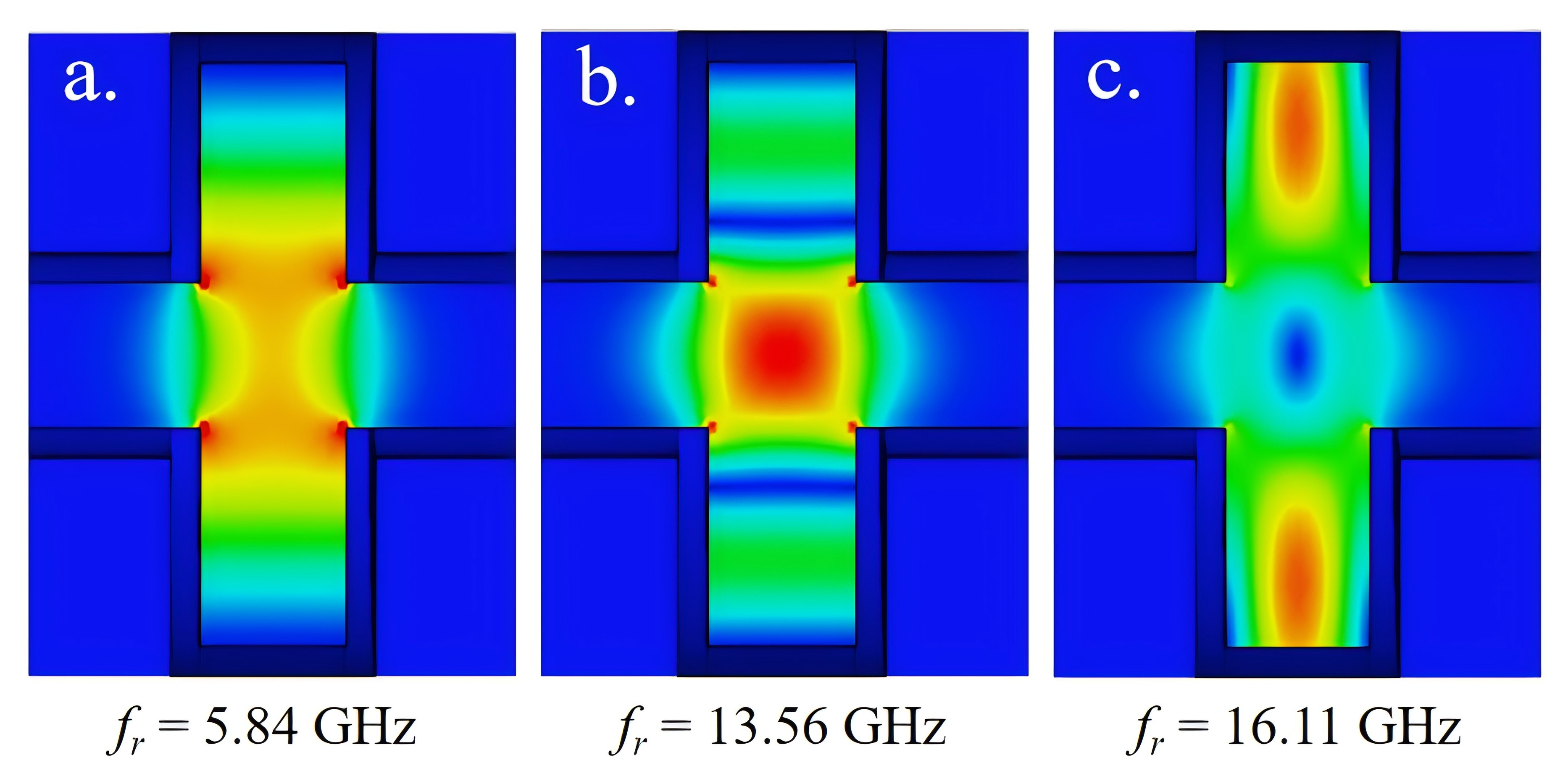}
    \caption{Electric field distribution of the trapped eigenmodes.}
    \label{fig:EigenMode3}
\end{figure}

To evaluate the accuracy of the fitting results, the pillbox cavity was analyzed using the CST eigenmode solver. Figure~\ref{fig:EigenMode3} shows the corresponding field distributions of the modes. The electric fields of the three modes are mainly localized in the cavity region, with no obvious propagation toward the beam pipes, indicating that these modes are trapped eigenmodes in the pillbox cavity.

Table~\ref{tab:long_param} compares the trapped-mode parameters obtained using the PSO method, the DE method, and the CST eigenmode solver. The results show good agreement among the three methods, especially for the resonant frequency $f_r$ and $R/Q$. Compared with the eigenmode-solver results, the relative errors of the shunt impedance $R_s$ obtained using the PSO and DE methods are within 1.5\% and 6\%, respectively, indicating the high accuracy of the PSO method.
\begin{table}[htbp]
    \centering
    \footnotesize
    \setlength{\tabcolsep}{4pt}
    \renewcommand{\arraystretch}{1.1}
    \caption{Comparison of longitudinal trapped-mode parameters of the pillbox cavity obtained using the PSO method, the DE method, and the CST eigenmode solver.}
    \label{tab:long_param}
    \begin{tabular*}{\linewidth}{@{\extracolsep{\fill}}
        c
        S[table-format=2.3] S[table-format=2.3] S[table-format=2.3]   
        S[table-format=2.2] S[table-format=2.2] S[table-format=2.2]   
        S[table-format=5.0] S[table-format=5.0] S[table-format=5.0]   
        S[table-format=3.1] S[table-format=3.1] S[table-format=3.1]   
        @{}}
    \toprule
    \multirow{2}{*}{Mode number} &
        \multicolumn{3}{c}{$f_r$ [GHz]} &
        \multicolumn{3}{c}{$R/Q$ [$\Omega$]} &
        \multicolumn{3}{c}{$Q$} &
        \multicolumn{3}{c}{$R_s$ [k$\Omega$]} \\
    \cmidrule(lr){2-4} \cmidrule(lr){5-7} \cmidrule(lr){8-10} \cmidrule(lr){11-13}
        & {PSO} & {DE} & {Eigen} & {PSO} & {DE} & {Eigen} & {PSO} & {DE} & {Eigen} & {PSO} & {DE} & {Eigen} \\
    \midrule
        1 &  5.844 &  5.844 &  5.843 & 66.53 & 66.89 & 66.49 &  7949 &  7475 &  7988 & 528.8 & 500.0 & 531.1 \\
        2 & 13.561 & 13.560 & 13.558 & 15.58 & 15.58 & 15.42 & 12330 & 12004 & 12640 & 192.1 & 187.0 & 194.9 \\
        3 & 16.108 & 16.110 & 16.108 &  1.99 &  1.99 &  1.96 &  7967 &  7792 &  7942 &  15.8 &  15.5 &  15.6 \\
    \bottomrule
    \end{tabular*}
\end{table}

\begin{figure}[htbp]
    \centering
    \includegraphics[width=0.5\linewidth]{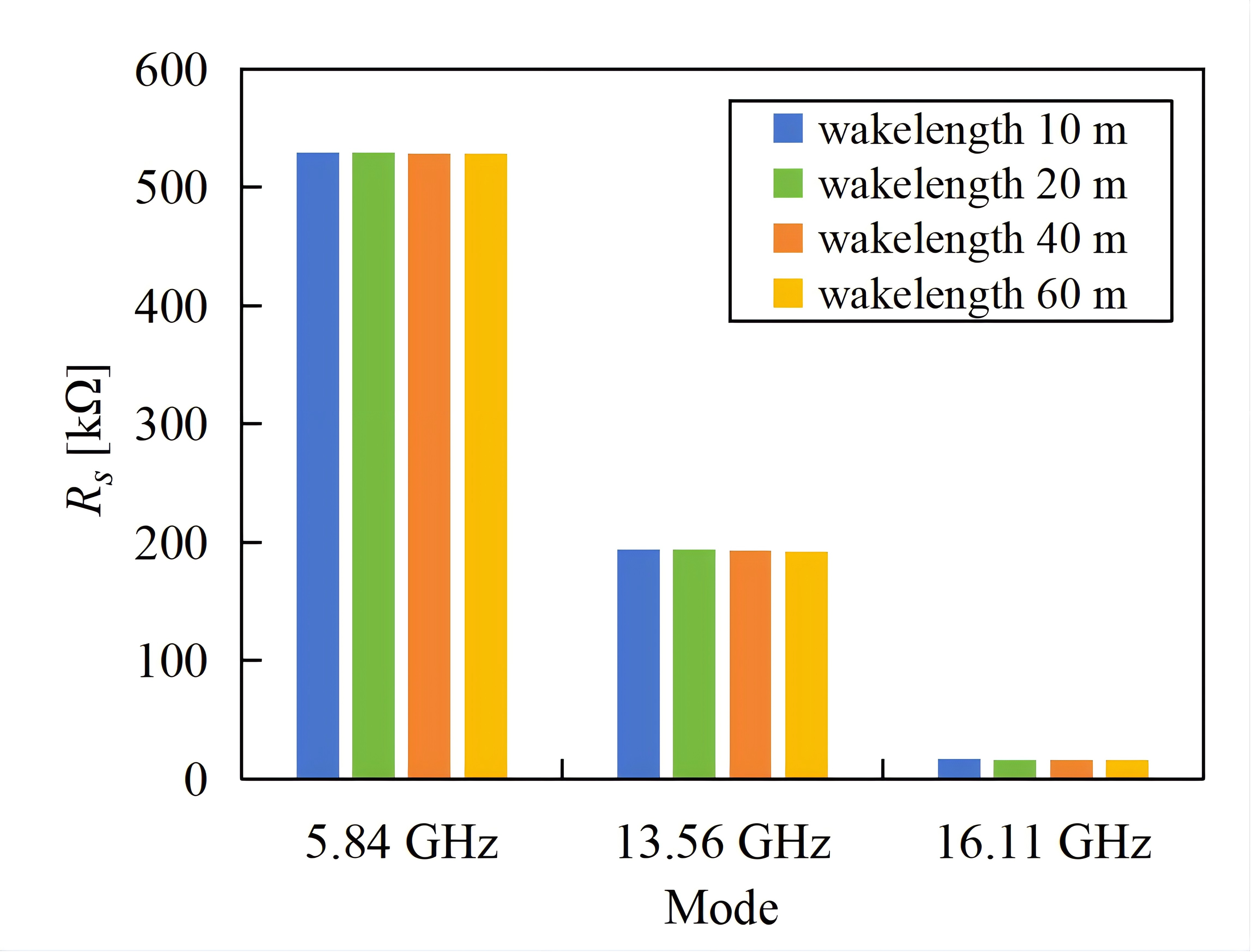}
    \caption{Comparison of reconstructed longitudinal shunt impedances obtained from PSO fitting using different wake potential lengths (10~m, 20~m, 40~m, and 60~m).}
    \label{fig:pillbox_shortwake}
\end{figure}

Further numerical results indicate that even when the wake potential length used for fitting is reduced to 40~m, 20~m, or even 10~m, the dominant trapped mode parameters can still be extracted in a stable and reliable manner. The corresponding reconstructed shunt impedance values are shown in Fig.~\ref{fig:pillbox_shortwake}. Compared with the results obtained from the 60~m wake-potential fitting, the impedance errors of all identified modes remain below 3\% when different wake lengths are used. These results demonstrate that the PSO method can accurately recover the trapped-mode parameters using truncated wake-potential data, thereby significantly reducing the computational cost of time-domain simulations while maintaining high fitting accuracy.

\subsection{Transverse trapped-mode parameter extraction results}
\label{sec:level3.2}
To further verify the applicability of the PSO method to the extraction of transverse trapped-mode parameters, a fitting analysis was performed on the transverse wake potential of the pillbox cavity. In the CST wakefield simulation, the transverse beam offset was set to \(2~\mathrm{mm}\), and the wake length used for fitting was \(20~\mathrm{m}\). Table~\ref{tab:trans_param} compares the transverse trapped-mode parameters obtained using three different methods. The results extracted by the PSO method agree well with the CST eigenmode values and the DE fitting results, with only minor deviations. Taking the CST eigenmode values as the reference, the maximum relative errors of \(R/Q\) obtained using the PSO and DE methods are within \(0.8\%\) and \(1.6\%\), respectively, while those of \(R_s\) are within \(4\%\). These discrepancies mainly arise from the finite propagation length of the wake potential and numerical fitting errors.
\begin{table}[htbp]
    \centering
    \footnotesize
    \setlength{\tabcolsep}{4pt}
    \renewcommand{\arraystretch}{1.1}
    \caption{Comparison of transverse trapped-mode parameters of the pillbox cavity obtained using the PSO method, the DE method, and the CST eigenmode solver.}
    \label{tab:trans_param}
    \begin{tabular*}{\linewidth}{@{\extracolsep{\fill}}
        c
        S[table-format=2.3] S[table-format=2.3] S[table-format=2.3]   
        S[table-format=4.0] S[table-format=4.0] S[table-format=4.0]   
        S[table-format=4.0] S[table-format=4.0] S[table-format=4.0]   
        S[table-format=2.2] S[table-format=2.2] S[table-format=2.2]   
        @{}}
    \toprule
    \multirow{2}{*}{Mode number} &
        \multicolumn{3}{c}{$f_{r}$ [GHz]} &
        \multicolumn{3}{c}{$R/Q$ [$\Omega$/m]} &
        \multicolumn{3}{c}{$Q$} &
        \multicolumn{3}{c}{$R_{s}$ [M$\Omega$/m]} \\
    \cmidrule(lr){2-4} \cmidrule(lr){5-7} \cmidrule(lr){8-10} \cmidrule(lr){11-13}
        & {PSO} & {DE} & {Eigen} & {PSO} & {DE} & {Eigen} & {PSO} & {DE} & {Eigen} & {PSO} & {DE} & {Eigen} \\
    \midrule
        1 & 8.933 & 8.930 & 8.932 & 6769 & 6775 & 6759 & 8479 & 8473 & 8560 & 57.40 & 57.40 & 57.85 \\
        2 & 14.069 & 14.070 & 14.074 &  863 &  870 &  857 & 4797 & 4575 & 4833 &  4.14 &  3.98 &  4.14 \\
        3 & 16.139 & 16.140 & 16.143 & 2078 & 2076 & 2079 & 6097 & 6070 & 6201 & 12.67 & 12.60 & 12.89 \\
        4 & 16.567 & 16.570 & 16.571 &  607 &  605 &  606 & 5883 & 6037 & 5801 &  3.57 &  3.65 &  3.52 \\
    \bottomrule
    \end{tabular*}
\end{table}

\subsection{Efficiency comparison}
\label{sec:level3.3}
The comparison of the longitudinal and transverse trapped-mode fitting results for the pillbox cavity shows that the PSO method achieves good fitting accuracy while offering a significant advantage in computational efficiency. Taking the longitudinal fitting as an example, under the same computational conditions, namely the same computer, the same wake-potential or impedance data, 800 iterations, and a population size of 500, the PSO method requires only about 10~s, whereas the DE method with the adaptive-bound strategy requires about 5~min. When the population size of the DE algorithm is reduced to 45 according to the commonly used empirical setting, approximately five times the optimization dimensionality, fitting the same problem still takes about 30~s, with a slight decrease in fitting accuracy. These results indicate that the PSO-based long-range wake fitting method has a clear advantage in computational efficiency.

\section{Application to critical components in the HALF storage ring}
The HALF is a 4th generation of storage ring light source under construction in Hefei, China. It comprises a full-energy linac, a transfer line, and a storage ring designed for a natural emittance below \(100~\mathrm{pm}\cdot\mathrm{rad}\) with 20 long straight sections and 20 short ones~\cite{Bai}. Complex vacuum components in such a ring can excite trapped modes and contribute significantly to beam coupling impedance. Based on the impedance modeling of HALF \cite{He2}, the gate valve, horizontal collimator, and in-vacuum undulator (IVU) are critical components that demand particular attention. These three components are selected as realistic examples to demonstrate the PSO-based extraction of trapped-mode parameters.

\subsection{Longitudinal impedance of gate valve}
Gate valves are essential components in accelerator vacuum systems, enabling isolation of beamline sections for vacuum pumping, leak detection, and maintenance by means of a movable gate. In the HALF storage ring, two gate valves are installed in each cell, resulting in approximately 40 units distributed around the entire ring. Owing to their cavity-like structure and large number, the impedance contribution of gate valves could be critical in beam stability analyses.

\begin{figure}[htbp]
    \centering
    \begin{subfigure}{0.8\textwidth}
        \centering
        \includegraphics[width=\linewidth]{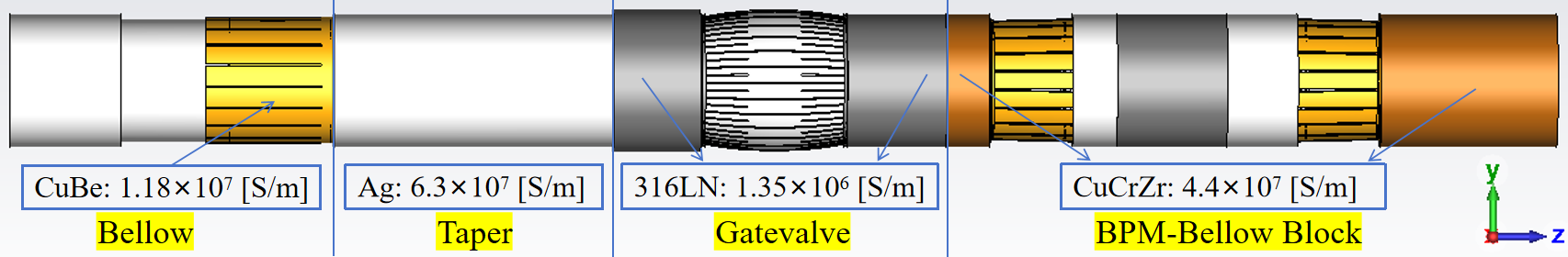}
        \caption{}
        \label{fig:gatevalve_model}
    \end{subfigure}

    \vspace{0.5em}

    \begin{subfigure}{0.8\textwidth}
        \centering
        \includegraphics[width=\linewidth]{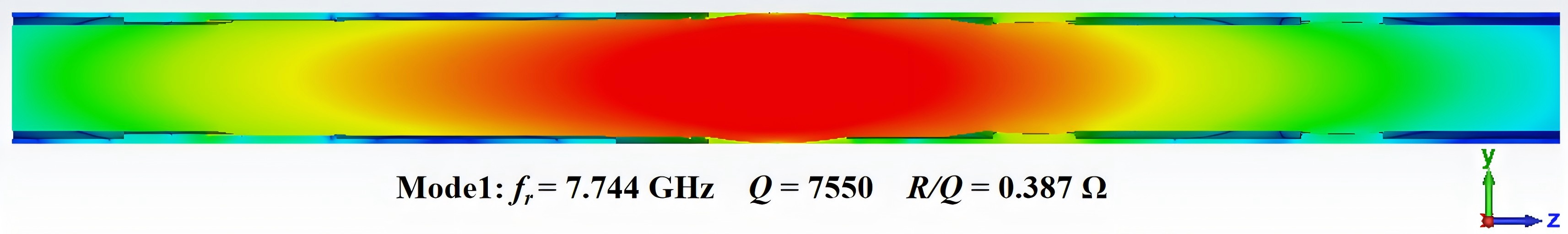}
        \caption{}
        \label{fig:gatevalve_eigenmode1}
    \end{subfigure}

    \caption{(a) Model of the gate valve with adjacent components, and (b) electric field distribution of the dominant trapped eigenmode.}
    \label{fig:gatevalve_model_field}
\end{figure}

Considering possible electromagnetic coupling between the gate valve and its nearby components such as an RF shielding bellows, a tapered transition, and a BPM-bellow block, the simplified integrated CST model is constructed, as shown in Fig.~\ref{fig:gatevalve_model_field}(a). The regions assigned silver material in the model are in fact stainless-steel substrates coated with a silver layer thicker than \(5~\mu\mathrm{m}\). Since this thickness exceeds the skin depth within the frequency range of interest, these regions are treated as silver in the electromagnetic modeling. Taking advantage of the component's axisymmetric structure, a quarter-symmetry model is adopted in the wake-potential simulations to reduce the computational cost. The bunch length is set to 5~mm, and the wake length is limited to 10~m, corresponding to a partially decayed wake potential.

\begin{figure}[htbp]
    \centering
    \begin{subfigure}{0.48\textwidth}
        \centering
        \includegraphics[width=\linewidth]{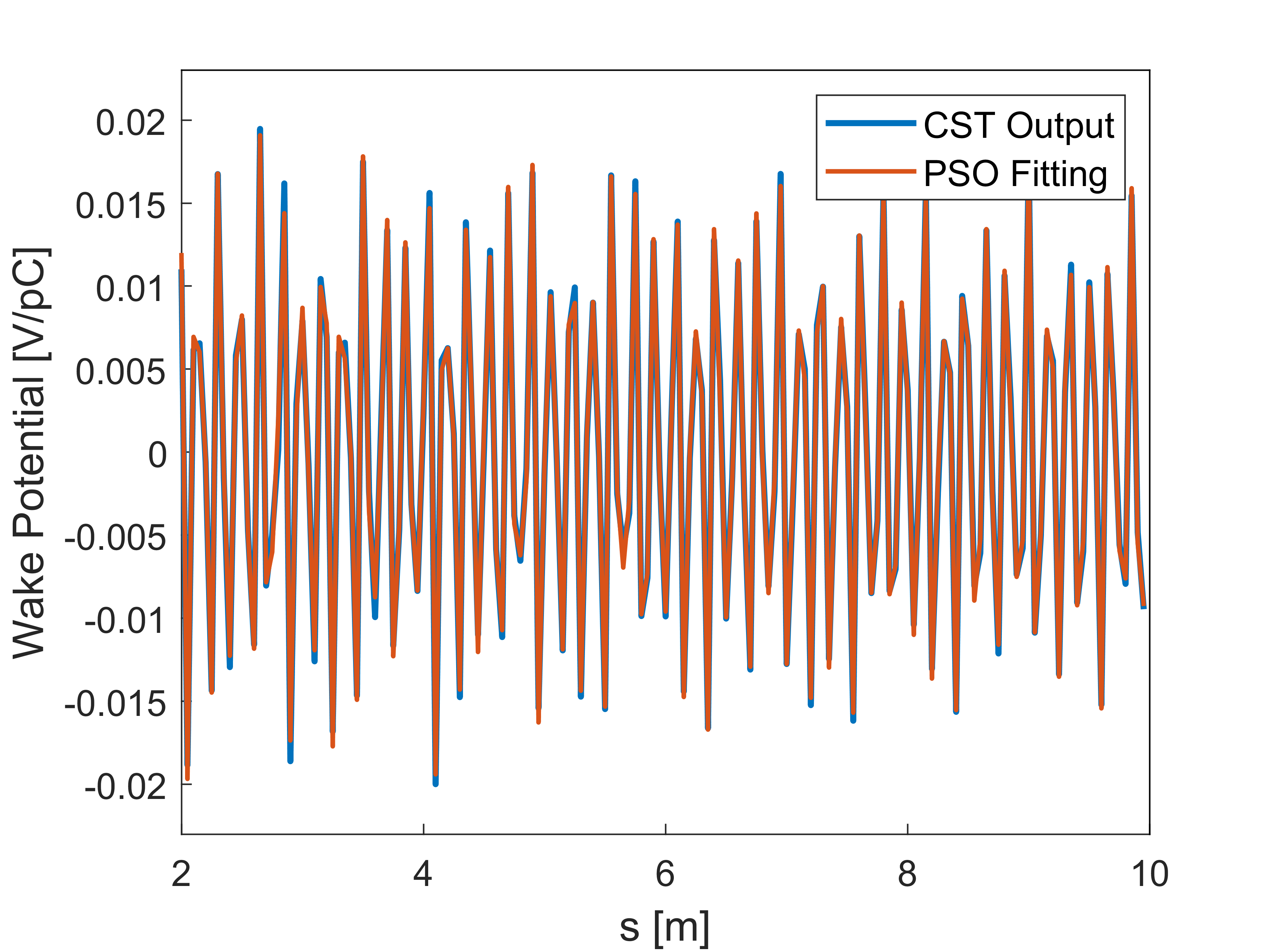}
        \caption{}
        \label{fig:gatevalve_wp}
    \end{subfigure}
    \hfill
    \begin{subfigure}{0.48\textwidth}
        \centering
        \includegraphics[width=\linewidth]{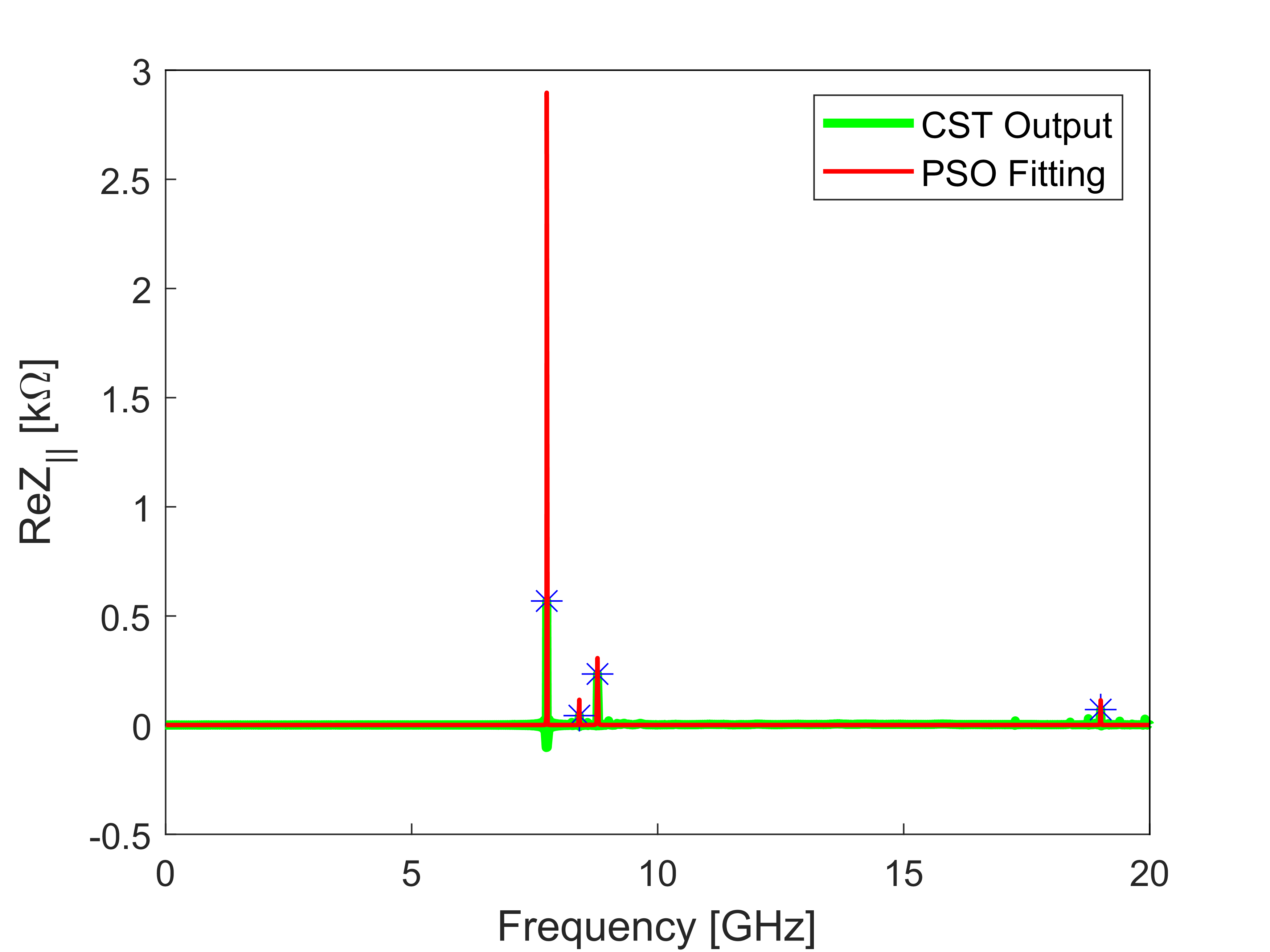}
        \caption{}
        \label{fig:gatevalve_imp}
    \end{subfigure}
    \caption{PSO-based results for the gate valve: (a) fitted longitudinal wake potential and (b) reconstructed longitudinal impedance.}
    \label{fig:gatevalve_results}
\end{figure}

Figure~\ref{fig:gatevalve_results} shows the fitted longitudinal wake potential and reconstructed impedance using the PSO method. The impedance obtained directly from the wakefield solver (with incomplete wake decay) significantly underestimates the dominant trapped mode, whereas the impedance reconstructed from PSO-fitted resonator parameters accurately recovers the true impedance.

Table~\ref{tab:gatevalve_long} compares the trapped-mode parameters extracted by the PSO method and the DE method. Four longitudinal trapped modes are identified. The resonant frequencies \(f_r\) and \(R/Q\) values obtained using the two methods agree well, whereas certain deviations are observed in \(Q\) and \(R_s\). In addition, the dominant Mode~1 was calculated using the CST eigenmode solver, yielding \(f_r=7.744~\mathrm{GHz}\), \(R/Q=0.387~\Omega\), and \(Q=7550\), corresponding to \(R_s=2.922~\mathrm{k}\Omega\). This mode exhibits a relatively high shunt impedance and represents one of the main longitudinal impedance contributors in the HALF storage ring. Taking the eigenmode-solver results as the reference, the relative errors of \(R/Q\) and \(R_s\) obtained using the PSO method are approximately 0.52\% and 0.85\%, respectively, while the corresponding errors for the DE method are approximately 1.03\% and 3.14\%. This result also indicates that the long-range wake fitting method can be used to evaluate the effects of electromagnetic crosstalk between adjacent components on impedance parameters, as well as impedance variations caused by local structural changes such as bellows-finger deformation.
\begin{table*}[htbp]
    \centering
    \footnotesize
    \setlength{\tabcolsep}{5pt}
    \renewcommand{\arraystretch}{1.1}
    \caption{Comparison of longitudinal trapped-mode parameters of the gate valve obtained using the PSO method and the DE method.}
    \label{tab:gatevalve_long}
    \begin{tabular*}{\linewidth}{@{\extracolsep{\fill}}
        c
        S[table-format=2.3] S[table-format=2.3]   
        S[table-format=1.3] S[table-format=1.3]   
        S[table-format=4.0] S[table-format=4.0]   
        S[table-format=1.3] S[table-format=1.3]   
        @{}}
    \toprule
    \multirow{2}{*}{Mode number} &
        \multicolumn{2}{c}{$f_r$ [GHz]} &
        \multicolumn{2}{c}{$R/Q$ [$\Omega$]} &
        \multicolumn{2}{c}{$Q$} &
        \multicolumn{2}{c}{$R_s$ [k$\Omega$]} \\
    \cmidrule(lr){2-3} \cmidrule(lr){4-5} \cmidrule(lr){6-7} \cmidrule(lr){8-9}
        & {PSO} & {DE} & {PSO} & {DE} & {PSO} & {DE} & {PSO} & {DE} \\
    \midrule
        1 &  7.746 &  7.746 & 0.389 & 0.391 & 7439 & 7246 & 2.897 & 2.830 \\
        2 &  8.409 &  8.408 & 0.029 & 0.030 & 3989 & 3301 & 0.115 & 0.100 \\
        3 &  8.778 &  8.778 & 0.242 & 0.245 & 1267 & 1243 & 0.306 & 0.304 \\
        4 & 19.002 & 19.000 & 0.027 & 0.029 & 4130 & 3619 & 0.114 & 0.105 \\
    \bottomrule
    \end{tabular*}
\end{table*}

The corresponding field distribution of Mode 1 is shown in Fig.~\ref{fig:gatevalve_model_field}(b). The electric field is mainly concentrated in the gate-valve region and gradually decreases along the beam-pipe direction on both sides, indicating that this trapped mode is mainly induced by the gate-valve structure. Meanwhile, the field distribution extends into the adjacent regions, suggesting that nearby components can affect the boundary conditions and resonant parameters of this mode. Therefore, it is necessary to use an integrated model including the adjacent components. Overall, the PSO method can reliably extract the longitudinal trapped-mode parameters of the gate valve from partially decayed wake-potential data.

\subsection{Transverse impedance of horizontal collimator}
\begin{figure}[htbp]
    \centering
    \begin{subfigure}[b]{0.33\textwidth}
        \centering
        \includegraphics[width=\textwidth]{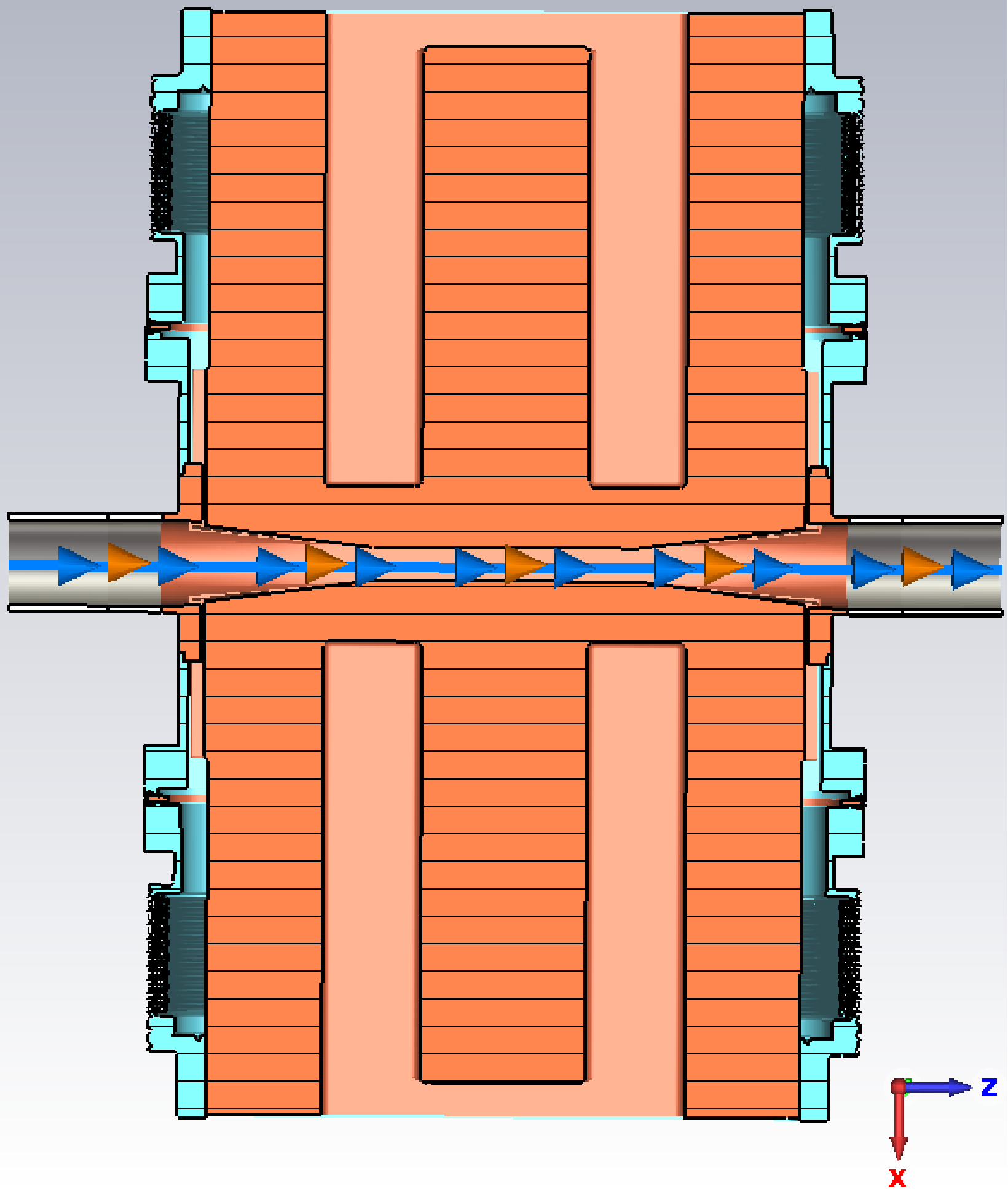}
        \caption{}
        \label{fig:collimator_model}
    \end{subfigure}
    \hfill
    \begin{subfigure}[b]{0.66\textwidth}
        \centering
        \includegraphics[width=\textwidth]{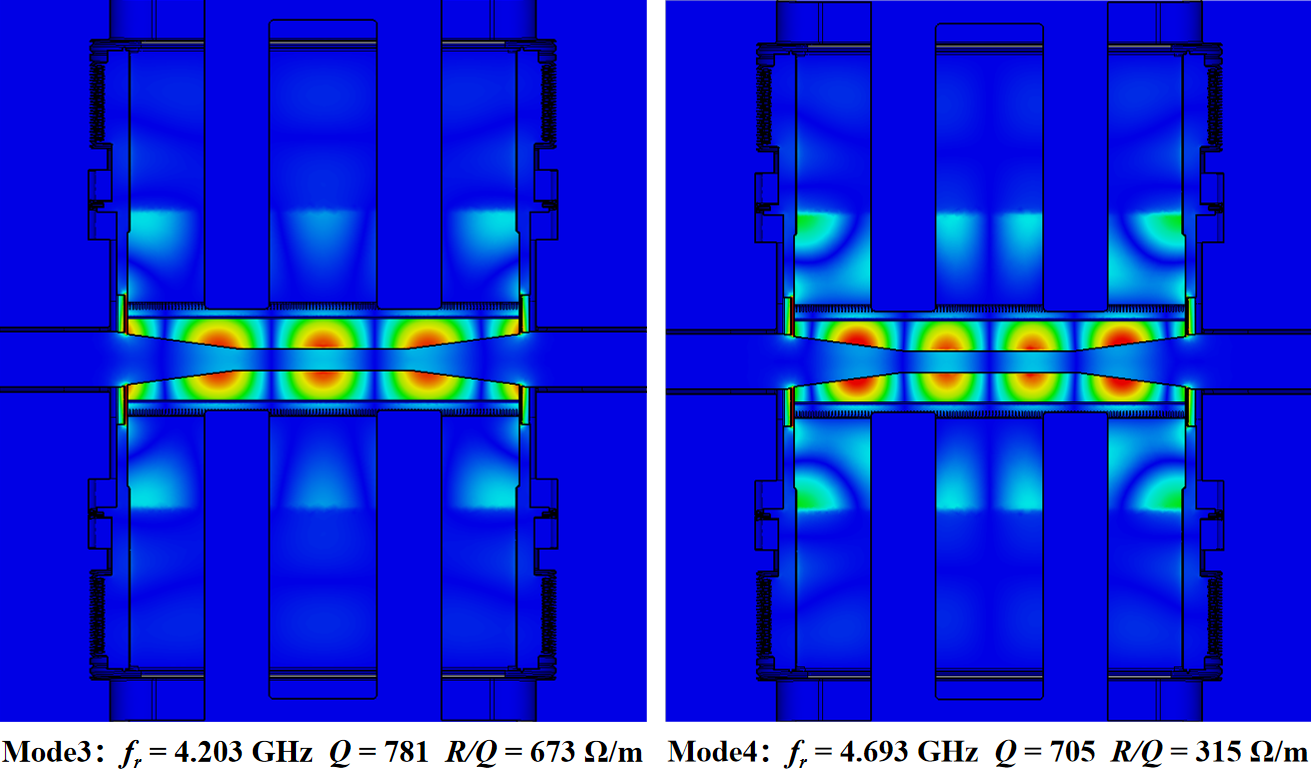}
        \caption{}
        \label{fig:collimator_eigen_mode3_4}
    \end{subfigure}
    \caption{(a) Horizontal collimator model and (b) electric field distributions of the two strongest transverse trapped eigenmodes.}
    \label{fig:collimator}
\end{figure}

Two identical horizontal collimators will be installed on the HALF storage ring, located in the high \(\beta_x\) region. Their main functions are to limit the transverse beam size, localize beam losses and radiation dose, protect insertion devices, and serve as a beam absorber and diagnostic tool~\cite{Drozhdin,Redaelli}. Their complex geometry and narrow aperture excite multiple horizontal dipole trapped modes, making them a challenging test case. A detailed CST model of the horizontal collimator is constructed based on the actual geometry, as shown in Fig.~\ref{fig:collimator}(a).

\begin{figure}[htbp]
    \centering
    \begin{subfigure}{0.49\textwidth}
        \centering
        \includegraphics[width=\linewidth]{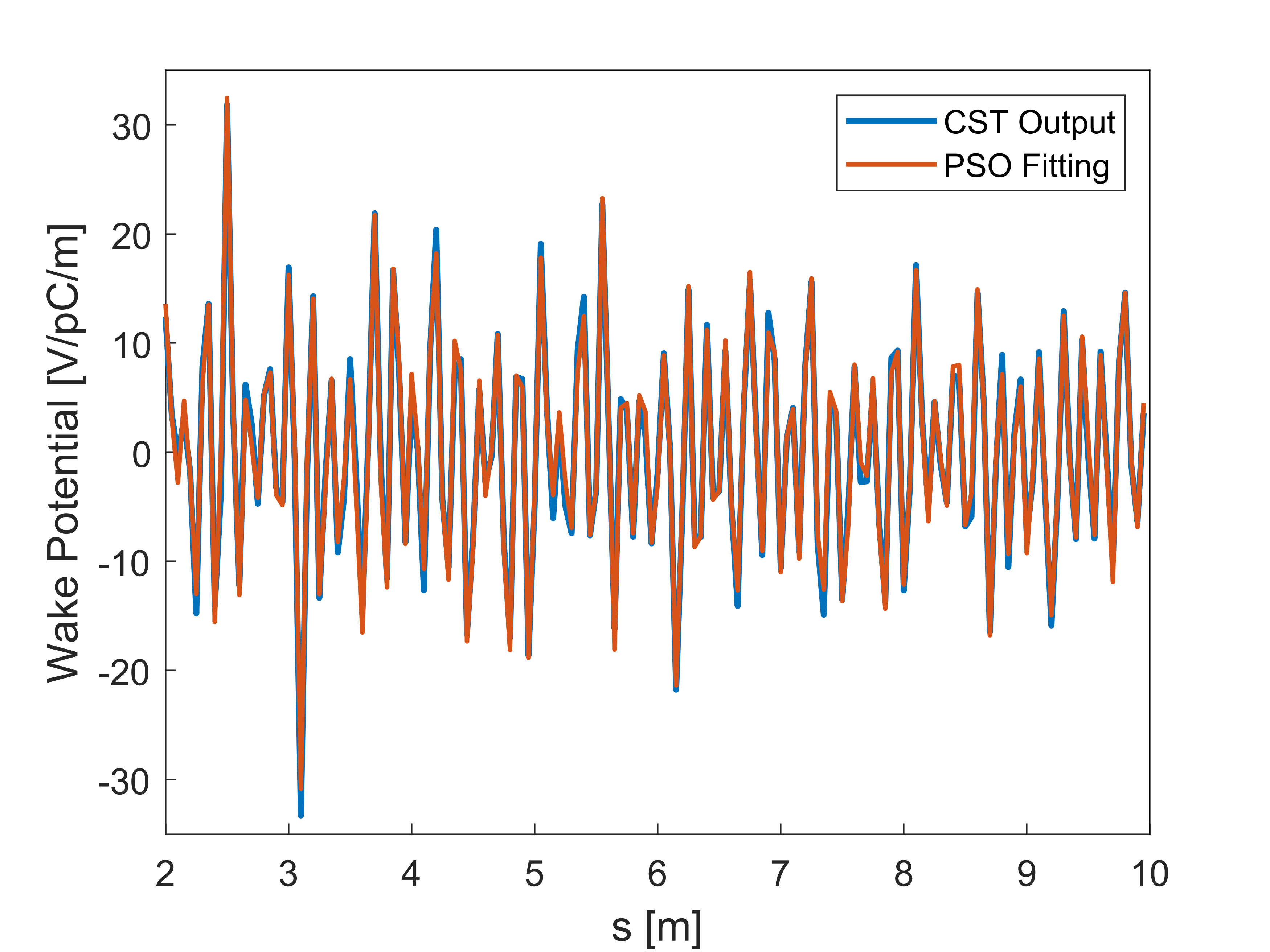}
        \caption{}
        \label{fig:collimator_wp}
    \end{subfigure}
    \hfill
    \begin{subfigure}{0.49\textwidth}
        \centering
        \includegraphics[width=\linewidth]{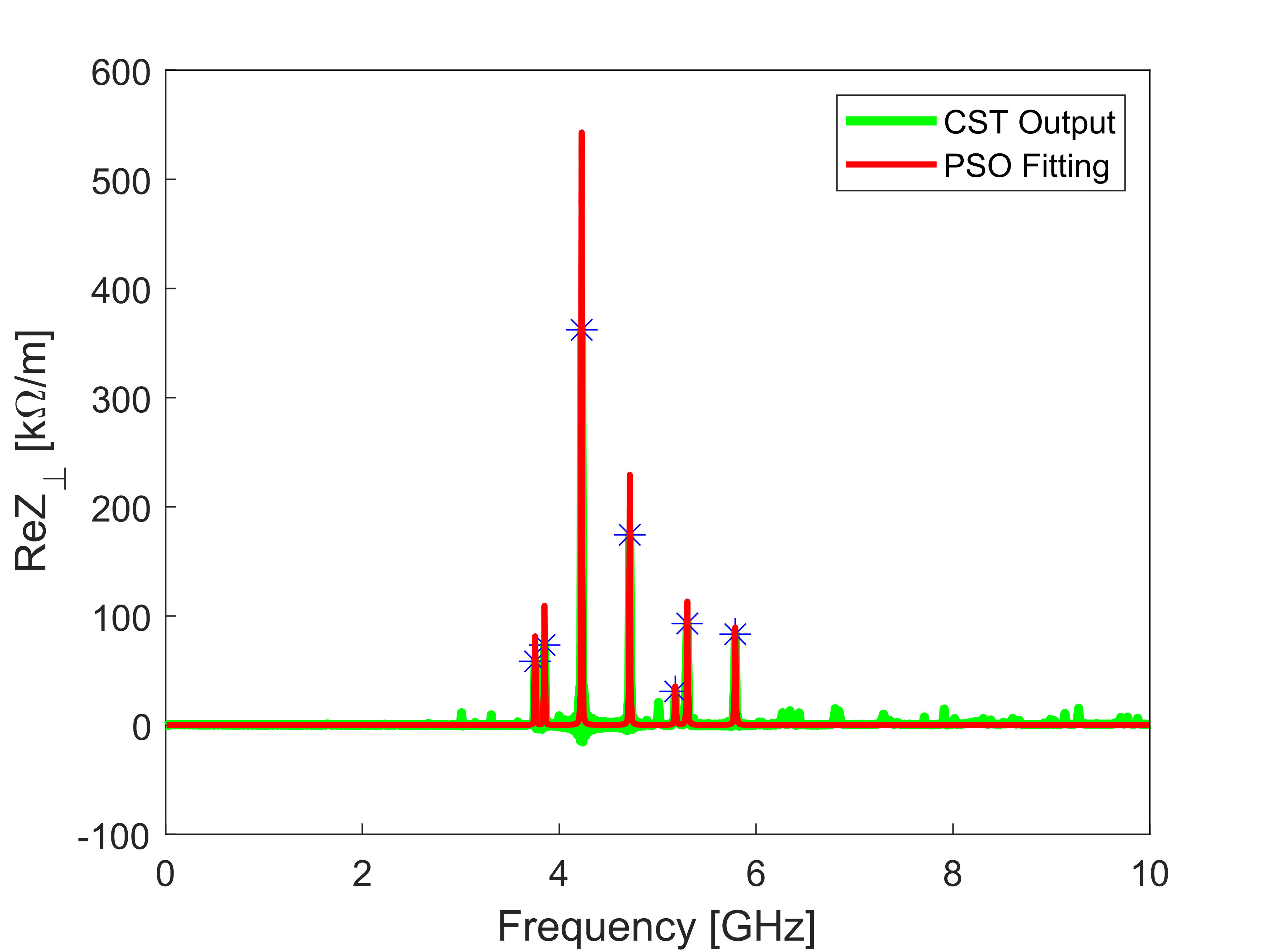}
        \caption{}
        \label{fig:collimator_imp}
    \end{subfigure}
    \caption{PSO-based results for the collimator: (a) fitted horizontal wake potential and (b) reconstructed horizontal impedance.}
    \label{fig:collimator_wp_imp}
\end{figure}

Since the horizontal collimator has mirror symmetry in both the vertical and horizontal planes, a quarter-symmetry model can be adopted in the CST wakefield solver to reduce the computational cost. To cover the frequency range of the main trapped modes, the bunch length is set to \(5~\mathrm{mm}\), the horizontal beam offset is set to \(1~\mathrm{mm}\), and the wake length is set to \(10~\mathrm{m}\). Based on the simulation results, modes with impedance peaks larger than \(30~\mathrm{k}\Omega/\mathrm{m}\) are selected, and their resonant frequencies are used as prior inputs for fitting the long-range horizontal wake potential. The fitted wake potential and reconstructed impedance spectrum are shown in Fig.~\ref{fig:collimator_wp_imp}(a) and (b), respectively. The wake fit nearly overlaps the reference data, indicating good fitting accuracy, while the reconstructed spectrum accurately reproduces the corresponding impedance peaks.

Table~\ref{tab:collimator_trans} presents the horizontal trapped-mode parameters extracted using the PSO-based method and DE-based method. Since multiple trapped modes are identified, the number of fitting variables increases accordingly, making the optimization process more time-consuming. With \(800\) iterations and the same number of fitting data points, the population size of the DE algorithm is set to \(105\), following the empirical rule that the population size is approximately five times the optimization dimensionality. Under this setting, the DE algorithm requires about \(166~\mathrm{s}\) to complete the fitting, whereas the PSO method requires only about \(7~\mathrm{s}\). It should be noted that the fitting accuracy decreases with this reduced population size. When the population size is increased to \(500\), the same value used in the preceding pillbox-cavity benchmark, the DE method requires about \(26~\mathrm{min}\), while the PSO method requires only about \(15~\mathrm{s}\). These results show that, as the number of trapped modes and the optimization dimensionality increase, the PSO-based long-range wake fitting method exhibits a more pronounced advantage in computational efficiency for multiple trapped-mode parameter extraction. The parameters obtained using the two optimization algorithms show certain differences, but the overall trends remain consistent. Among them, Mode~3 and Mode~4 have relatively large shunt impedances and therefore represent important transverse geometric impedance contributions. 
\begin{table*}[htbp]
    \centering
    \footnotesize
    \setlength{\tabcolsep}{5pt}
    \renewcommand{\arraystretch}{1.1}
    \caption{Comparison of horizontal trapped-mode parameters of the horizontal collimator obtained using the PSO-based method and DE-based method.}
    \label{tab:collimator_trans}
    \begin{tabular*}{\linewidth}{@{\extracolsep{\fill}}
        c
        S[table-format=1.3] S[table-format=1.3]   
        S[table-format=3.1] S[table-format=3.1]   
        S[table-format=3.1] S[table-format=3.1]   
        S[table-format=3.1] S[table-format=3.1]   
        @{}}
    \toprule
    \multirow{2}{*}{Mode number} &
        \multicolumn{2}{c}{$f_r$ [GHz]} &
        \multicolumn{2}{c}{$R/Q$ [$\Omega$/m]} &
        \multicolumn{2}{c}{$Q$} &
        \multicolumn{2}{c}{$R_s$ [k$\Omega$/m]} \\
    \cmidrule(lr){2-3} \cmidrule(lr){4-5} \cmidrule(lr){6-7} \cmidrule(lr){8-9}
        & {PSO} & {DE} & {PSO} & {DE} & {PSO} & {DE} & {PSO} & {DE} \\
    \midrule
        1 & 3.755 & 3.755 & 125.6 & 128.5 & 646.4 & 664.4 &  81.2 &  85.4 \\
        2 & 3.851 & 3.851 & 144.8 & 139.6 & 754.1 & 795.3 & 109.2 & 111.0 \\
        3 & 4.228 & 4.228 & 671.1 & 664.0 & 809.0 & 838.9 & 543.0 & 557.0 \\
        4 & 4.717 & 4.717 & 324.9 & 318.5 & 705.4 & 740.9 & 229.2 & 236.0 \\
        5 & 5.179 & 5.179 &  63.9 &  75.3 & 550.8 & 430.1 &  35.2 &  32.4 \\
        6 & 5.302 & 5.301 & 176.2 & 178.9 & 641.4 & 609.2 & 113.0 & 109.0 \\
        7 & 5.789 & 5.789 & 141.5 & 131.5 & 631.0 & 783.4 &  89.3 & 103.0 \\
    \bottomrule
    \end{tabular*}
\end{table*}

Due to the complex geometry of the collimator, eigenmode calculations require a relatively high mesh resolution, and a complete eigenmode analysis of all trapped modes would be computationally expensive. Therefore, only Mode~3 and Mode~4 are verified using the CST eigenmode solver. For Mode~3, the eigenmode calculation gives \(f_r=4.203~\mathrm{GHz}\), \(R/Q=673~\Omega/\mathrm{m}\), and \(Q=781\), corresponding to \(R_s=525.6~\mathrm{k}\Omega/\mathrm{m}\). For Mode~4, the corresponding values are \(f_r=4.693~\mathrm{GHz}\), \(R/Q=315~\Omega/\mathrm{m}\), and \(Q=705\), corresponding to \(R_s=222.1~\mathrm{k}\Omega/\mathrm{m}\). Taking the eigenmode results as the reference, the maximum relative errors of \(R/Q\) obtained using the PSO and DE methods are approximately 3.14\% and 1.34\%, respectively; the corresponding maximum relative errors of \(R_s\) are approximately 3.31\% and 6.27\%. These errors are within an acceptable range, indicating that the dominant trapped-mode parameters obtained using the two fitting methods remain reliable. The corresponding field distributions are shown in Fig.~\ref{fig:collimator}(b). The electric fields of Mode~3 and Mode~4 are mainly concentrated near the beam pipe, indicating that these two transverse trapped modes originate primarily from the gap structures near the beam pipe and the localized electromagnetic boundaries formed by the complex collimator geometry.

The above results demonstrate that, even for geometrically complex accelerator components, the PSO method can stably and accurately extract the dominant transverse trapped-mode parameters of the horizontal collimator from finite-length wake-potential data, while requiring a lower computational cost.

\subsection{Transverse impedance of the short IVU}
\begin{figure}[htbp]
    \centering
    \begin{subfigure}{0.5\textwidth}
        \centering
        \includegraphics[width=\linewidth]{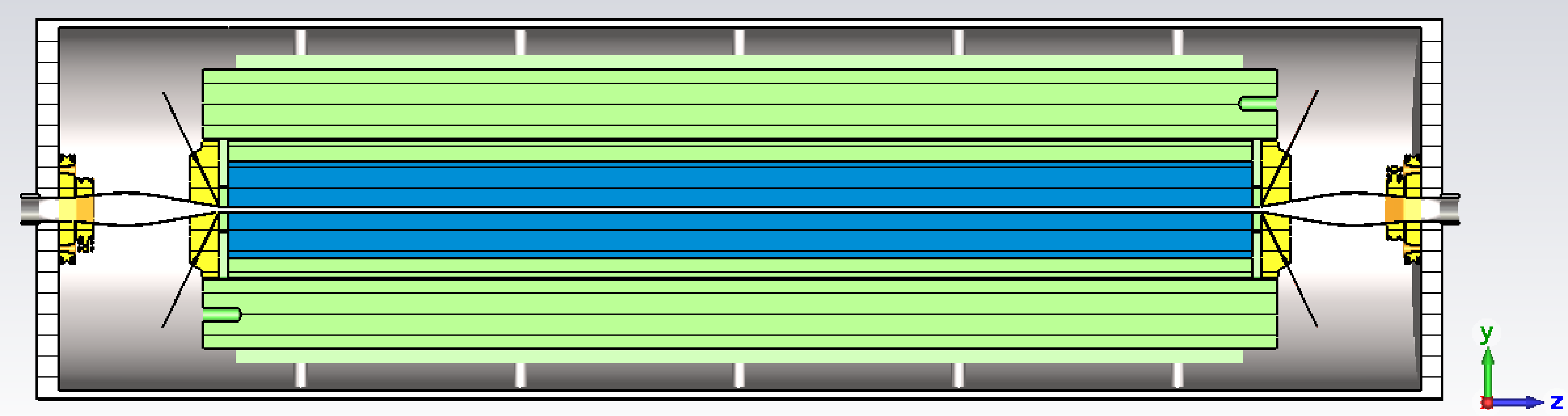}
        \caption{}
        \label{fig:Short_IVU_model}
    \end{subfigure}

    \vspace{0.5em}

    \begin{subfigure}{0.5\textwidth}
        \centering
        \includegraphics[width=\linewidth]{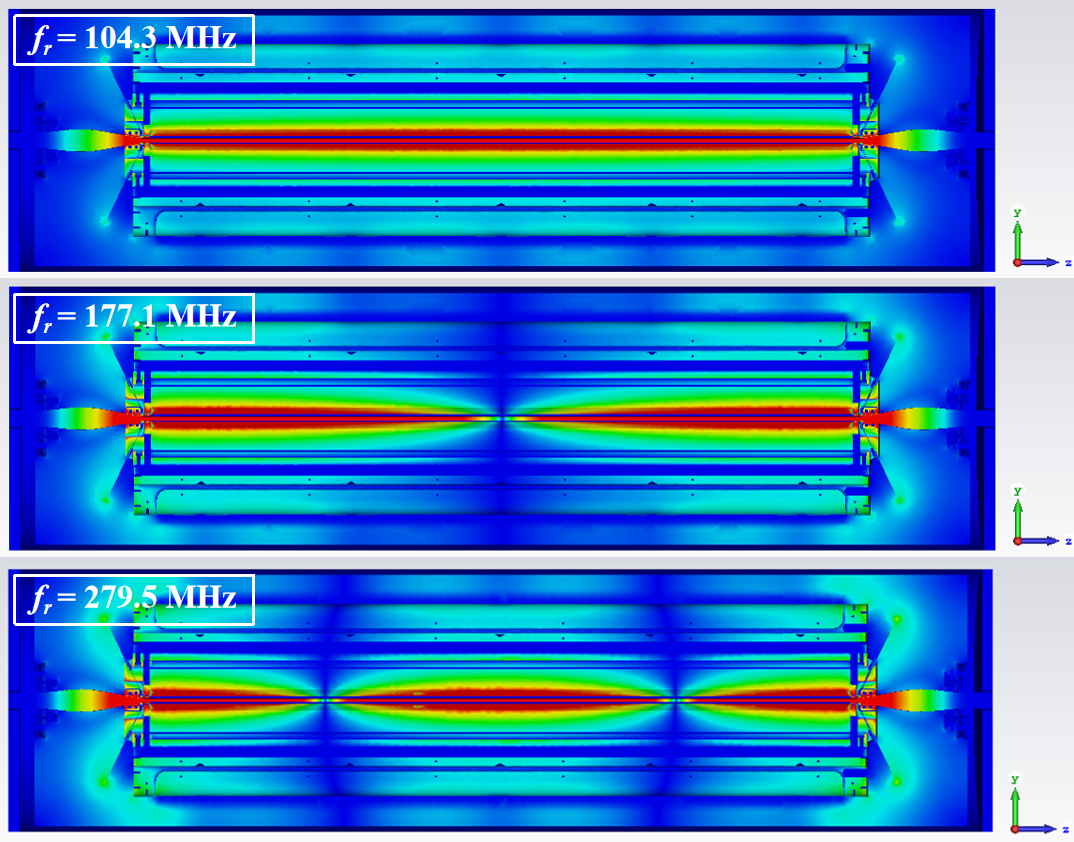}
        \caption{}
        \label{fig:Short_IVU_eigenmode}
    \end{subfigure}

    \caption{(a) Short IVU model and (b) electric field distributions of the three dominant trapped eigenmodes.}
    \label{fig:Short_IVU_model_field}
\end{figure}

The IVU can generate high-quality photon beams with high photon energy, higher flux, and greater brightness. However, its extremely narrow magnetic gap and ridge-waveguide-like structure may excite strong trapped-mode impedance in the vertical plane, which can drive transverse coupled-bunch instabilities~\cite{Tian,Dowd,Wang}. The HALF storage ring contains one long IVU and one short IVU. In this work, the short IVU is selected as a case study, and its CST model is shown in Fig.~\ref{fig:Short_IVU_model_field}(a). Since this component is relatively long and has complex geometrical details, wakefield simulations of the complete model require a high computational cost. Therefore, a simplified model is adopted in this study, in which structures such as pumping ports and water pipes are neglected. It should be noted that these simplifications affect the quantitative evaluation of the actual impedance characteristics of the IVU~\cite{Yao}. Therefore, the results presented here are used only to verify the feasibility of the proposed method and should not be regarded as the actual impedance of the HALF IVU.

\begin{figure}[htbp]
    \centering
    \begin{subfigure}{0.48\textwidth}
        \centering
        \includegraphics[width=\linewidth]{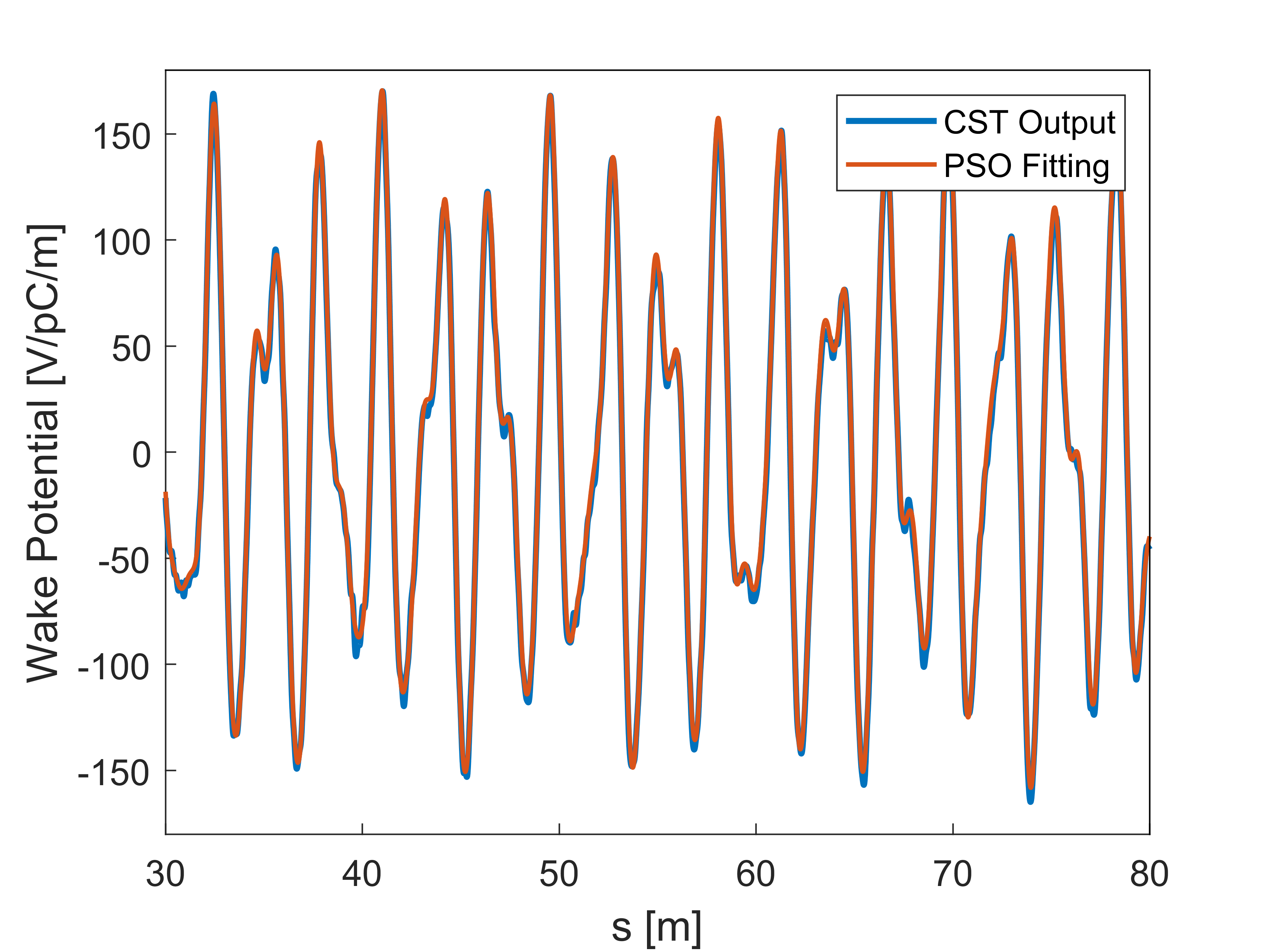}
        \caption{}
        \label{fig:IVU_wp}
    \end{subfigure}
    \hfill
    \begin{subfigure}{0.48\textwidth}
        \centering
        \includegraphics[width=\linewidth]{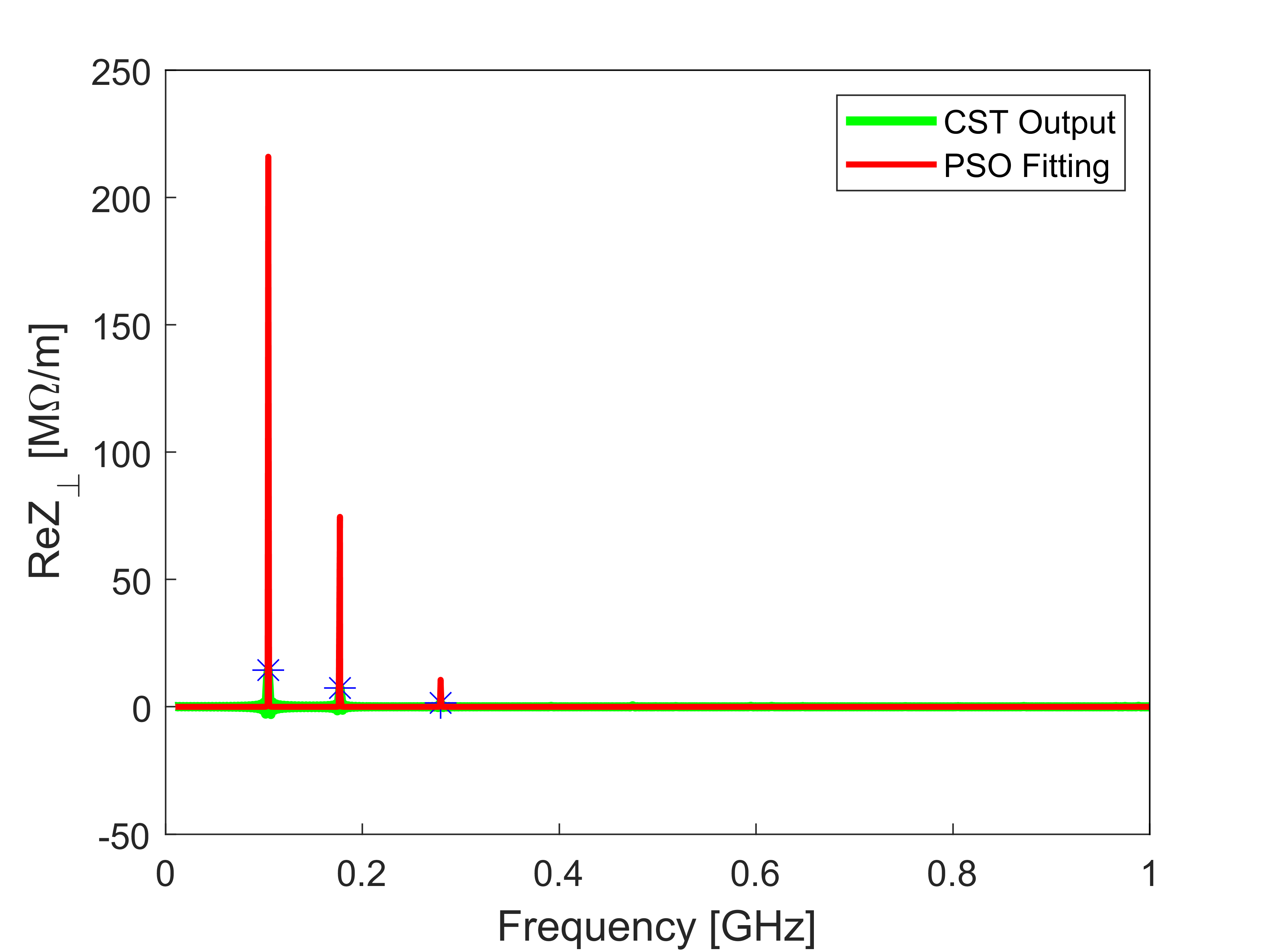}
        \caption{}
        \label{fig:IVU_imp}
    \end{subfigure}
    \caption{PSO-based results for the short IVU: (a) fitted vertical wake potential and (b) reconstructed vertical impedance.}
    \label{fig:IVU_wp_imp}
\end{figure}

Since the dominant trapped modes of the short IVU are concentrated in the hundreds of megahertz range, to fully cover the critical frequency range, the bunch length in the CST wakefield solver is set to \(50~\mathrm{mm}\), which corresponds to an effective frequency range of approximately \(2~\mathrm{GHz}\). Meanwhile, the vertical beam offset is set to \(1~\mathrm{mm}\) and the wake length is set to \(80~\mathrm{m}\). Figure~\ref{fig:IVU_wp_imp} shows the fitting result of the vertical wake potential of the IVU obtained using the PSO method, together with the vertical impedance spectrum reconstructed from the fitted parameters. The fitted wake potential agrees well with the CST simulation result, and the reconstructed impedance indicates that the short IVU exhibits extremely strong narrowband trapped-mode impedance in the vertical direction. Therefore, it is one of the transverse impedance sources that should be carefully considered in the HALF storage ring.

Table~\ref{tab:IVU_trans} summarizes the transverse trapped-mode parameters of the short IVU extracted using different methods. Taking the eigenmode results as the reference, for the strongest trapped mode, the relative deviations of \(R_s\) are approximately \(1.1\%\) for the PSO method and \(5.4\%\) for the DE method, while the corresponding deviations of \(R/Q\) are approximately \(3.2\%\) and \(2.8\%\). These results indicate that, even for IVUs with large dimensions and complex geometries, the proposed long-range wake fitting method can effectively extract their dominant trapped-mode parameters.
\begin{table*}[htbp]
    \centering
    \footnotesize
    \setlength{\tabcolsep}{5pt}
    \renewcommand{\arraystretch}{1.1}
    \caption{Comparison of vertical trapped-mode parameters of the short IVU obtained using the PSO method, the DE method, and the CST eigenmode solver.}
    \label{tab:IVU_trans}
    \begin{tabular*}{\linewidth}{@{\extracolsep{\fill}}
        c
        S[table-format=3.1] S[table-format=3.1] S[table-format=3.1]   
        S[table-format=3.1] S[table-format=3.1] S[table-format=3.1]   
        S[table-format=4.0] S[table-format=4.0] S[table-format=4.0]   
        S[table-format=3.1] S[table-format=3.1] S[table-format=3.1]   
        @{}}
    \toprule
    \multirow{2}{*}{Mode number} &
        \multicolumn{3}{c}{$f_r$ [MHz]} &
        \multicolumn{3}{c}{$R/Q$ [k$\Omega$/m]} &
        \multicolumn{3}{c}{$Q$} &
        \multicolumn{3}{c}{$R_s$ [M$\Omega$/m]} \\
    \cmidrule(lr){2-4} \cmidrule(lr){5-7} \cmidrule(lr){8-10} \cmidrule(lr){11-13}
        & {PSO} & {DE} & {Eigen} & {PSO} & {DE} & {Eigen} & {PSO} & {DE} & {Eigen} & {PSO} & {DE} & {Eigen} \\
    \midrule
        1 & 104.4 & 104.3 & 104.3 & 169.9 & 169.3 & 164.7 & 1271 & 1329 & 1296 & 215.9 & 225.0 & 213.5 \\
        2 & 177.2 & 177.2 & 177.1 &  52.1 &  52.0 &  50.3 & 1433 & 1462 & 1407 &  74.7 &  76.0 &  70.7 \\
        3 & 279.5 & 279.4 & 279.5 &   7.1 &   7.1 &   7.3 & 1490 & 1530 & 1657 &  10.6 &  10.8 &  12.1 \\
    \bottomrule
    \end{tabular*}
\end{table*}

The above results show that the short IVU has strong vertical trapped-mode impedance and is a major transverse impedance source in the HALF storage ring. Therefore, further impedance optimization based on a realistic model is necessary. Figure~\ref{fig:Short_IVU_model_field}(b) shows the eigenmode field distributions of the three main trapped modes. The electric fields of these modes are mainly localized in the narrow magnetic gap region and the end transition sections of the IVU, indicating that these regions are the main sources of the strong vertical trapped-mode impedance. This provides a useful reference for subsequent structural optimization, such as improving the end transition geometry or introducing suitable damping structures. The long-range wake fitting method can provide effective support for IVU impedance calculations and optimization assessment.

\section{Conclusion}\label{sec:level4} 
This paper presented a method for extracting narrowband impedance parameters based on PSO fitting of long-range wake potentials, aiming at the analysis of trapped modes in accelerator vacuum components. Validation on a pillbox cavity demonstrates that the PSO method can reliably extract trapped-mode parameters from partially decayed long-range wake potentials, with results that agree well with those obtained from the CST eigenmode solver and the DE method. The PSO method was then successfully applied to three critical vacuum components in the HALF storage ring. The results confirm that the PSO method is suitable for both longitudinal and transverse trapped-mode impedance analysis of realistic complex structures. In particular, for high-dimensional parameter extraction problems involving multiple trapped modes, the PSO method significantly improves computational efficiency while preserving fitting accuracy. These findings highlight the potential of the PSO fitting method for trapped-mode optimization of large, complex impedance components, such as IVUs and RF cavities.

The PSO method was applied to trapped-mode impedance analysis of all components in the HALF storage ring~\cite{He2}. A graphical user interface (GUI) has been developed to integrate data loading, parameter extraction, and result visualization into a unified platform, thereby improving the usability of the method. The GUI is available from the first corresponding author upon reasonable request.

\section{Acknowledgment}
The authors thank Tao Huang and Anxin Wang for providing the relevant CAD models. This work was supported by the National Natural Science Foundation of China (No. 12375324) and the Fundamental Research Funds for the Central Universities (No. WK2310000127).

\end{document}